\documentclass{tip}
\usepackage[title]{appendix}
\newcommand{\uproman}[1]{\uppercase\expandafter{\romannumeral#1}}
\newcommand{\autora}{R. Zöllner}
\newcommand{\autorb}{B. Kämpfer}
\title{\vspace{-2.5cm}\hrulefill\\\color{darkblue}\sffamily\LARGE\bfseries Core-Corona Decomposition of Very Compact (Neutron) Stars: 
	Accounting for Current Data of XTE J1814-338\\}
\author{\sffamily\bfseries \autora\orcidlink{0000-0002-3544-6622}${}^a$~\&~\autorb${}^{b,c}$%	~\&~\autorc
	}
\affil{\sffamily${}^a$Institut f\"ur Technische Logistik und Arbeitssysteme, TU~Dresden,\newline01062 Dresden, Germany, rico.zoellner@tu-dresden.de\\
\sffamily${}^{b}$Helmholtz-Zentrum  Dresden-Rossendorf, 01314 Dresden, Germany\\
\sffamily${}^{c}$Institut f\"ur Theoretische Physik, TU~Dresden, 01062 Dresden, Germany}
\date{\sffamily\today\\\hrulefill}

\begin{document}
	\maketitle
	\thispagestyle{plain}
	\vspace{-2.0em}
	
	\parbox[t]{0.40\linewidth}{\textbf{Keywords}
		\begin{itemize}
			\item compact stars 
			\item core-corona~decomposition 
			\item impact of core mass
			\item dark matter admixture
			\end{itemize}}
	\hfill
	\parbox[t]{0.55\linewidth}{\textbf{Abstract}\\A core-corona decomposition of compact (neutron) star models was compared with the current mass-radius data of the outlier XTE~J1814-338.
	The corona (which may also be dubbed the envelope, halo or outer crust) is assumed to be of Standard Model matter, with an equation of state that is supposed to be faithfully known and accommodates nearly all other neutron star data. The core, solely parameterized by its mass, radius and transition pressure, presents a challenge regarding its composition. We derived a range of core parameters needed to describe the current data of XTE J1814-338.}
	
	\vspace{2.0em} 
	\noindent\rule{\textwidth}{0.4pt}
	\vspace{-3.0em}
	
\section{Introduction} \label{introduction}
The {pulse-profile} modeling of the {accretion-powered} millisecond pulsar XTE J1814-338 by a single {uniform-temperature}
{hot-spot} resulted in mass ($M$) and radius ($R$) parameters {of}
$(R/\mathrm{km}, M/M_\odot)$ = ($7.0^{+0.4}_{-0.4}$,  $1.21^{+0.05}_{-0.05}$) \cite{Kini:2024ggu}.
In particular, the~extraordinarily small radius poses a problem for establishing a proper neutron star model
based on a unique equation of state (EoS) {that} is compatible with other compact (neutron) star data.
The latter ones are provided by
gravitational waves from merging neutron stars,
the related {multi-messenger} astrophysics 
and the improving {mass-radius} determinations, in~particular {those} by NICER data {(see Appendix A in~\cite{Zollner:2023myk} for a survey and relationship to other data).}
Canonical neutron star radii are centered at about $12~\mathrm{km}$.
Indeed, ref.~\cite{Miller:2021qha}  quotes $R =12.45^{+0.65}_{-0.65}~\mathrm{km}$ for a $1.4M_\odot$ neutron star and 
$R = 12.35^{+0.75}_{-0.75}~\mathrm{km}$ for a $2.08M_\odot$ neutron star.
HESS J1731-347  with ($10.4^{+0.86}_{-0.78}$,  $0.77^{+0.20}_{-0.17}$) \cite{Doroshenko:2022nwp} may be another outlier with respect to mass\footnote{The very small mass value is scrutinized in~\cite{Salmi:2024bss}.},
as the black widow pulsar PSR J0952-0607 may also be, which points to a large mass of $2.35^{+0.17}_{-0.17}~M_\odot$~\cite{Romani:2022jhd}, with implications studied in~\cite{Ecker:2022dlg}.\\
The problem of XTE J1814-338 is addressed in~\cite{Pitz:2024xvh,Yang:2024ycl,Laskos-Patkos:2024fdp,Zollner:2024ufa,Lopes:2025jyz,Lopes:2024ixl}.
Various proposals have been forwarded thereby, including various forms of Dark Matter (DM) admixtures and 
a strong {first-order} phase transition in Standard Model (SM) matter
at {supra-nuclear} {densities} (see also~\cite{Li:2022ivt} for an avenue {toward} {ultra-compact} hybrid stars).
Scenarios with DM components offer a multitude of {set-ups}, e.g.,\ including
fermionic or bosonic or mirror matter in various~constellations.\\
Instead of adding another specific model of the matter composition to explain XTE J1814-338 in terms of {the mass} and radius,
{here we followed} an agnostic approach, outlined in~\cite{Zollner:2022dst,Zollner:2023myk,Zollner:2024ufa},
known as {the core-corona} decomposition (CCD).
This approach can be imagined as a subdivision of a compact (neutron) star into two parts -- the core, embedded in the corona\footnote{Our notion 
		``corona'' is a synonym for  {``mantel'', ``crust'', ``envelope'', ``shell'' or ``halo''}.
		It refers to the complete part of the compact star outside the core, {where} $r_x \le r \le R$.}. The core is solely parameterized by its {radius, mass} and surface  pressure. The~latter one supports
	the corona,{which, by~definition, is} composed {of} SM {matter only,} with {a} known equation of
	state (EoS). Imposing {a} hydrodynamic equilibrium of the isotropic corona fluid, the~{Einsteinian}
	equations for spherical symmetry {completely govern} the corona structure. Either uncertainties of the {supra-nuclear} EoS or unknown DM admixtures (or other exotic {non-SM} matter
	effects) or both are not explicitly dealt with but {are} completely condensed into the three core parameters. The~motivation of this approach is {isolated and quantifies lesser} or uncertainly
	known features of matter in the deep interior of a compact star with the hope of receiving
	hints {about} their nature. If~a core is not required or one meets a ``trivial~core'' (this notion is
	explained below), the~conventional approach to spherically symmetric {static and} compact stars
	is recovered.\\
Our note is organized as follows.
Section~\ref{sect:II} outlines the CCD.
A comparison of the CCD with data is presented in Section~\ref{sect:III},
where we include XTE J1814-338 {as a compact} object with a large massive core
(which may contain a DM component or/and a special SM material component).We also put emphasis on the mapping of the credibility regions of masses and radii
to the inferred core masses and radii.
Section~\ref{app:B} provides a brief comment on two particular core models, one
based on a statistically determined EoS from {multi-messenger} data
and another one accommodating a strong {first-order} phase transition.
We summarize in Section~\ref{sect:summary}. 
Appendix \ref{app:C} deals with an example where no core is required.
For the sake of {self-containment}, a~few results from~\cite{Zollner:2024ufa}
are recollected, where an extensive bibliography can be found. In Appendix~\ref{AppB}, we apply our CCD with a scale-free corona EoS. 
\section{Outline of {Core-Corona~Decomposition}}\label{sect:II}
The standard modeling approach of compact star configurations is based on 
the {Tolman-Oppenheimer-Volkoff} (TOV) equations
\begin{eqnarray}
	\frac{\mbox{d} p(r)}{\mbox{d} r } &=& - G_N \frac{[e(r)+p(r)] [m(r)+4~\pi~r^3 p(r)]}{r^2 [1 - 2 G_N \frac{m(r)}{r}]}, \label{eq:p_prime} \\
	\frac{\mbox{d} m(r)}{\mbox{d} r} &=& 4\pi r^2 e(r), \label{eq:m_prime}
\end{eqnarray}
resulting from the {energy-momentum} tensor of a static isotropic fluid 
(described locally by pressure $p$ and energy density $e$, {which are} solely relevant for the medium) 
and {the} spherical symmetry of both {space-time} and {matter} 
within the framework of {Einsteinian} gravity without {the} cosmological term~\cite{Schaffner-Bielich:2020psc}.
Newton's constant is denoted by $G_N$, and~natural units with $c = 1$ are used,
unless when relating mass, length, pressure and energy density, where $\hbar c$ is~needed.\\
Given a unique relationship between {the} pressure $p$ and {the} energy density $e$ {in the} EoS
$e(p)$, in~particular at zero temperature, the~TOV equations are {customarily integrated} {(e.g., with~the Runge-Kutta algorithm)} with boundary conditions
$p(r) = p_c - {\cal O} (r^2)$ and $m(r) = 0 + {\cal O} (r^3)$ at small radii $r$,
and $p(R) = 0$ and $m(R) = M${, with} $R$ as {the} circumferential radius and $M$ as {the} gravitational mass
(acting as {parameters} in the external (vacuum) Schwarzschild solution at $r > R$).
The quantity $p_c$ is the central pressure.
The solutions $R(p_c)$ and $M(p_c)$
provide the {mass-radius relationship} in {the} parametric form $M(R)$, being a~curve.\\
Here, we {employed} another approach~\cite{Zollner:2022dst,Zollner:2023myk,Zollner:2024ufa}.
We {parameterized} the {supra-nuclear} core by a radius $r_x$ and the
included mass $m_x$ and {integrated} the above TOV equations only within the corona,
i.e.,\ from pressure $p_x$
to the surface, where $p = 0$. This {yielded} the total mass $M(r_x, m_x; p_x)$ and the total radius $R(r_x, m_x; p_x)$
by assuming that the corona EoS $p(e)$ {was} reliably known for $p \le p_x$
and only SM matter {occupied} the region $r \ge r_x$. 
Clearly, without~any knowledge
of the matter composition at $p > p_x$ (may it be SM matter with an uncertainly known EoS or
may it contain a DM admixture, for~instance, or~ monopoles or some other type of
``exotic'' matter){, one does not obtain} a simple {mass-radius relationship} by such a procedure, 
but admissible area(s) over the {mass-radius} plane, depending on the core
parameters $r_x$ and $m_x$ and the matching pressure $p_x$ and related energy density $e_x$\footnote{There are various possibilities to constrain the ($p_x$, $m_x$, $r_x$) parameter space, e.g.,~fixing $p_x$ as the pressure
		at the nuclear saturation density -- thus defining the ``core'' as a part with {supra-nuclear} density, if~the above
		{one-fluid} TOV equations are supposed to hold, or~selecting the radius $r_x$ as the locus where the pressure
		of a DM admixture vanishes, meaning that the core is to be dealt with in {two-fluid} TOV equations. Other
		side conditions are conceivable, e.g.,~employing the same value of $p_x$ for several compact (neutron) stars,
		such as HESS J1731-347 and XTE J1814-338 considered in~\cite{Zollner:2024ufa}. Here, we do not impose constraints and
		assume the applicability of the EoS NY$\Delta$ in the corona up to the maximum pressure and energy density
		tabulated in~\cite{Li:2019fqe} and study the range of core parameters delivering the XTE J1814-338 mass and radius.}.
This is the price of avoiding a specified model of the core matter composition.
However, the~CCD is a simple and efficient approach to quantify 
the appearance of some ``exotics''
by a displacement from the {mass-radius} curve related to an SM matter EoS. 
For an SM matter-only core with EoS $p^\mathrm{SM} (e)$, the~core parameters are strongly correlated,
$m_x(p_c)$$, r_x(p_c)$ for $p_c \ge p_x$, thus yielding masses and radii near to or on the
$M(R)$ curve provided by  $p^\mathrm{SM} (e)$.
The limits $r_x \to 0$, $m_x \to 0$ and $p_x \to p_c$ reduce the CCD
in the conventional {one-fluid} TOV~equations.\\
Note that our {core-corona} approach relies on the assumption that
the region $r \in [r_x, R]$ is occupied only by SM matter with a trustable EoS.
Thus, scenarios as in~\cite{Pitz:2024xvh}, where bosonic DM forms a halo around
a core with SM + DM components, are not captured by our~CCD.
\newpage
\section{Catching Current Data of  XTE J1814-338 by~CCD}\label{sect:III}
\subsection{Selecting a Reference~EoS}\label{subsect:data}
The NICER data of  PSR J0437-4715~\cite{Choudhury:2024xbk},
$(R/\mathrm{km}, M/M_\odot)$ =
($11.36^{+0.95}_{-0.63}$,   $1.42^{+0.04}_{-0.04}$)
and PSR J0740+6620~\cite{Salmi:2024aum},
($12.49^{+1.28}_{-0.88}$, $2.073^{+0.069}_{-0.069}$),
exhibited in Figure~1 in~\cite{Zollner:2024ufa}, clearly demonstrate that the EoS NY$\Delta$\footnote{Actually, we combine the EoS N {(purely nucleonic matter based on DD-ME2 functional described in~\cite{Li:2019fqe})}
	for \mbox{$(p_1, e_1) < (8.19, 216.92)~\mathrm{MeV/fm}^3$} with NY$\Delta$ {(hyperson-$\Delta$-excitation admixed matter)} for {$\left(p_{\mathrm{max}}= 3.73.58~\mathrm{MeV/fm}^3,\right.$} {\mbox{$\left. e_{\mathrm{max}} = 1232.85~\mathrm{MeV/fm}^3\right)  > $}} $(p, e) > (p_1, e_1)$.
	At $(p, e) < (0.41, 68.2)~\mathrm{MeV/fm}^3$, we use a linear interpolation to $(p =0, e_0)$ with
	$e_0 = 1~\mathrm{MeV/fm}^3$. Analog linear interpolations apply for the tabulated values in 
	{Table \uproman{1}} in~\cite{Li:2019fqe}, {where further details can be found}.}
in~\cite{Li:2019fqe} satisfactorily matches these data.
However, it falls short in the maximum mass of 
the black widow pulsar PSR J0952-0607, which {points} to a large mass of $2.35^{+0.17}_{-0.17}~M_\odot$ 
\cite{Romani:2022jhd} with implications studies in~\cite{Ecker:2022dlg}.
More (NICER) data {are} mentioned in Appendix~A
in~\cite{Zollner:2024ufa} to exhibit 
the usefulness of NY$\Delta$, which, however, fails badly for the outlier XTE J1814-338~\cite{Kini:2024ggu}
and {still shows} some tension with ($10.4^{+0.86}_{-0.78}$,  $0.77^{+0.20}_{-0.17}$)  
for HESS J1731-347~\cite{Doroshenko:2022nwp}, which may be another outlier or a hint of twin stars~\cite{Li:2024sft}. %EE: Please check that intended meaning has been retained.
We {focus} here on a description of  XTE J1814-338, $(R/\mathrm{km}, M/M_\odot)$ =
($7.0^{+0.4}_{-0.4}$,  $1.21^{+0.05}_{-0.05}$) \cite{Kini:2024ggu}, by~CCD. 
Rotational effects {are} assumed to be~subleading.
\begin{figure}[H]
	\centering
	\includegraphics[width=0.69\columnwidth]{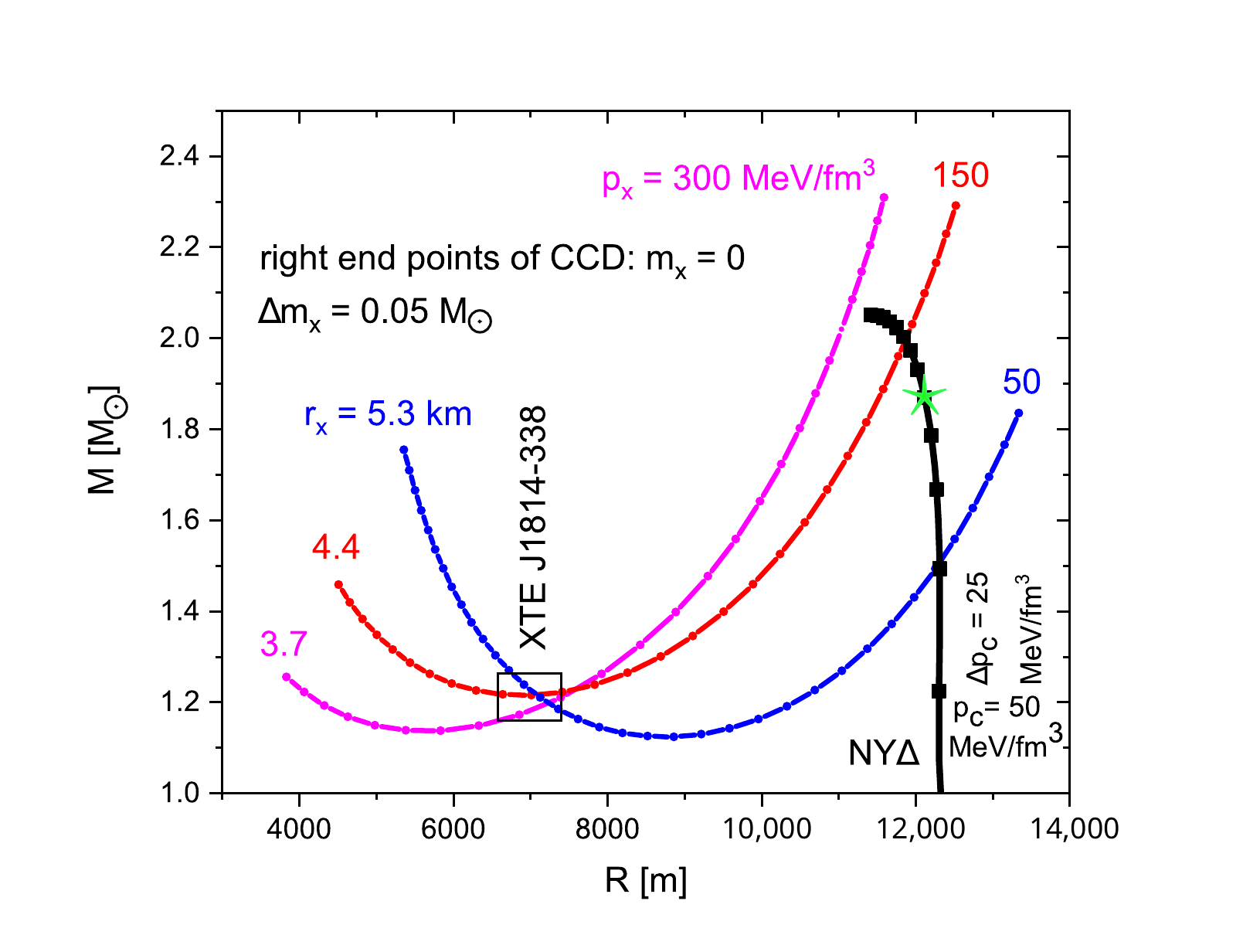} \hspace{-9mm}
	\\[-6mm]
	\caption{
		{{The mass-radius} relationship} of compact (neutron) stars in the CCD with core parameters
		$(  r_x/\mathrm{km}, p_x/[\mathrm{MeV/fm}^3])  = \left(  3.7, 300\right) $ (magenta curve),
		$\left(  4.4, 150\right) $ (red curve) and  $\left(  5.3, 50 \right) $ (blue curve).
		The right end points use $m_x = 0$; the dots depict increasing values of $m_x$ with an increment of
		$\Delta m_x = 0.05~M_\odot$ in going  left.  
		The black XTE square is for $\bar s = 1$.
		The fat solid black curve is for {no-core one-fluid} TOV equations with EoS NY$\Delta$. 
		The filled squares depict the loci of  $p_c = 25 n~\mathrm{MeV/fm}^3$ for $n = 2, \ldots, 14$. 
		The green asterisk exposes the point with $p_c = 150~\mathrm{MeV/fm}^3$.
		\label{fig:1} 
	}
\end{figure}
\subsection{Masses and~Radii}\label{masses}
\begin{figure}[H]
	\centering
	\includegraphics[width=0.6\columnwidth]{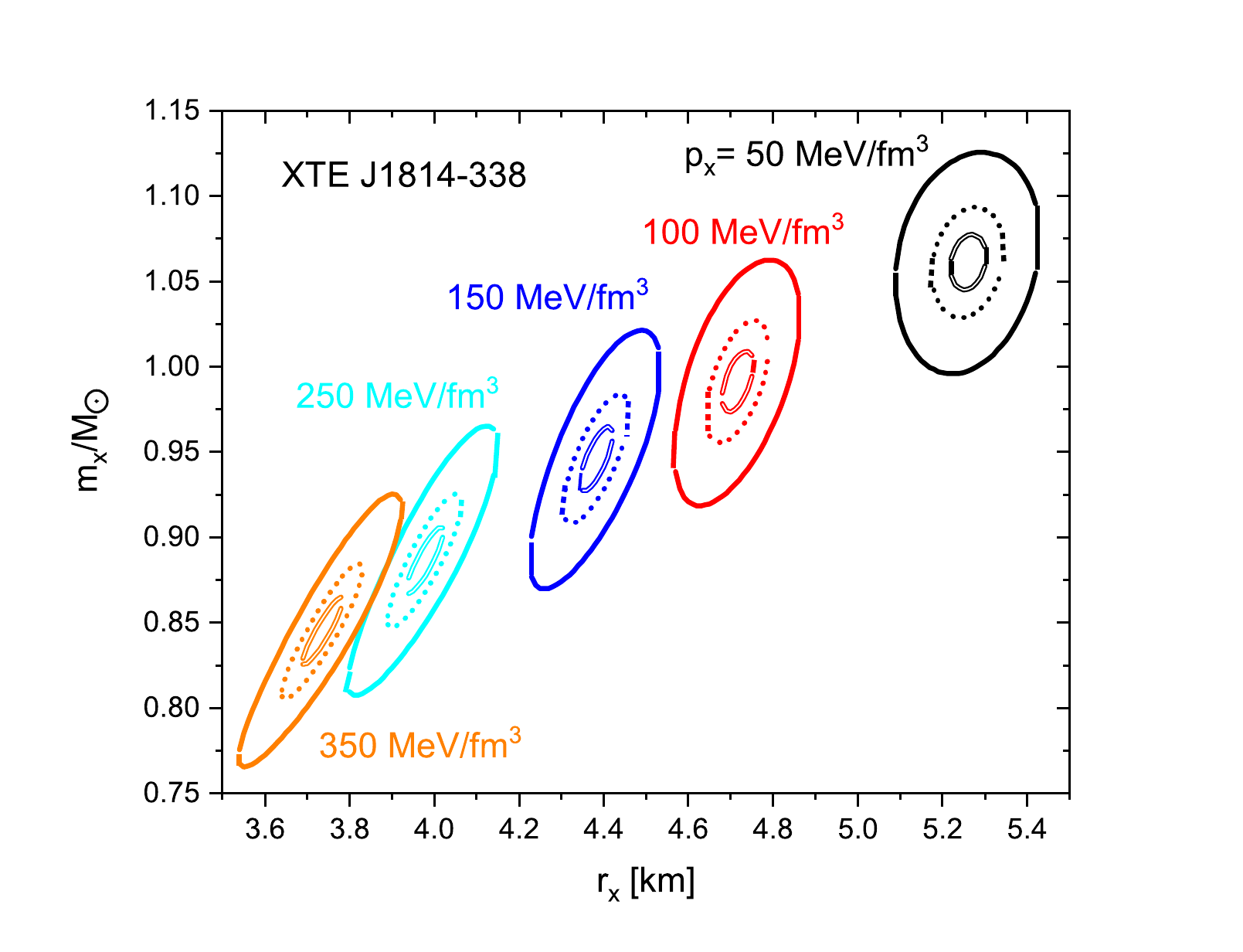} 
	\\[-6mm]
	\caption{
		Admissible areas of core masses $m_x$ and core radii $r_x$
		for $p_x = 50$ (black), 100 (red), 150~(blue), 250 (cyan) and $350~\mathrm{MeV/fm}^3$ (orange).
		The values of $m_x$ and $r_x$ are determined to deliver, at~a given $p_x$, masses $M$ and radii $R$
		within the XTE ellipses for $s = 1$ (solid curves), 2 (dotted curves) and 4 (double lines).
		\label{fig:1b} 
	}
\end{figure}

The credibility region of XTE J1814-338 {was} according to~\cite{Kini:2024ggu}
mass $M \in M_\mathrm{XTE} \pm \Delta_\mathrm{XTE}^{(M)} \bar s$ and 
radius $R \in R_\mathrm{XTE} \pm \Delta_\mathrm{XTE}^{(R)} \bar s$
with 
$M_\mathrm{XTE} = 1.21M_\odot$, $\Delta_\mathrm{XTE}^{(M)} = 0.05M_\odot$, 
$R_\mathrm{XTE} = 7.0~\mathrm{km}$, $\Delta_\mathrm{XTE}^{(R)} = 0.4~\mathrm{km}$
and $\bar s = 1$. We {named} this region ``XTE square''. Occasionally, we also {considered}
the ``XTE ellipsis'':
\begin{equation}
	\left(\frac{s}{\Delta_\mathrm{XTE}^{(M)}} \right)^2 (M - M_\mathrm{XTE})^2 + \left( \frac{s}{\Delta_\mathrm{XTE}^{(R)}}\right)^2  (R - R_\mathrm{XTE})^2 =1.
\end{equation}
In both cases, the~parameters $\bar s$ and $s$ {are} meant to steer an assumed accuracy
or to describe the credibility~level.\\
Selecting {core-surface} pressures $p_x = 50, 150$ and $300~\mathrm{MeV/fm}^3$ and core radii
$r_x = 5.3$ (blue curve), $4.4$ (red curve) and $3.7~\mathrm{km}$ (magenta curve), one {finds} good descriptions
of the XTE square for suitable values of core masses; see Figure~\ref{fig:1}. 
The curves {were} generated by running values of $m_x$ from zero (right end points) to larger values by steps of 
$\Delta m_x = 0.05~M_\odot$ in going left. %EE: Please check that intended meaning has been retained.  
In addition, the~{mass-radius relationship} for NY$\Delta$ without any core {is} exhibited by the solid black curve.
Filled squares are for $p_c = 25 n~\mathrm{MeV/fm}^3$ for $n = 2, \ldots, 14$. 
The asterisk exposes the point with $p_c = 150~\mathrm{MeV/fm}^3$.\\
{By replacing} the XTE square with $\bar s = 1$ with XTE ellipses, one {finds} the regions of core masses $m_x$ and
core radii $r_x$ displayed in Figure~\ref{fig:1b} for a sequence of selected values of $p_x = 50, 100, 150, 250$
and $350~\mathrm{MeV/fm}^3$. The~parameters {were set to} $s = 1$ (solid curves), 2 (dotted curves) and 4 (double lines).
The centers of the displayed ellipses approximately {obeyed} $m_x \approx (0.33 + 0.14 \frac{r_x}{\mathrm{km}}) M_\odot$,
which {allowed} space for $m_x  \approx 0.33~M_\odot$ when {extrapolated} to $r_x \to 0$. %EE: Please check that intended meaning has been retained. 
That is, the~core compactness $\mathcal{C}_x = 2 G_N m_x/r_x \approx 0.42 + \frac{\mathrm{km}}{r_x}$ 
{increased} with increasing {core-boundary} pressure $p_x$.
For instance, $\mathcal{C}_x\vert_{350~\mathrm{MeV/fm}^3}\approx 0.7$
{was} much larger than the total XTE compactness of 0.52. 
The core masses {scaled} with $p_x$ as $m_x \approx \left(1.513 - \log(p_x/[\mathrm{MeV/fm}^3]) \right) M_\odot$.
\noindent
Replacing the credibility ellipses with squares, the~ellipses turned into parallelograms with
envelopes approximately obeying
$m_x \approx (a + b r_x )~M_\odot$ with
$a = 0.262$, $b = 0.1366/\mathrm{km}$ (lower tips)
and $a = 0.3266$, $b = 0.1566/\mathrm{km}$ (upper tips), which leaves space for
\mbox{$m_x \in [0.262, 0.3266]~M_\odot$} when extrapolated to $r_x \to 0$.
\subsection{Pressure and Mass Profiles in the~Corona}\label{subsect:profiles}
\begin{figure}[H]
	\centering
	\includegraphics[width=0.49\columnwidth]{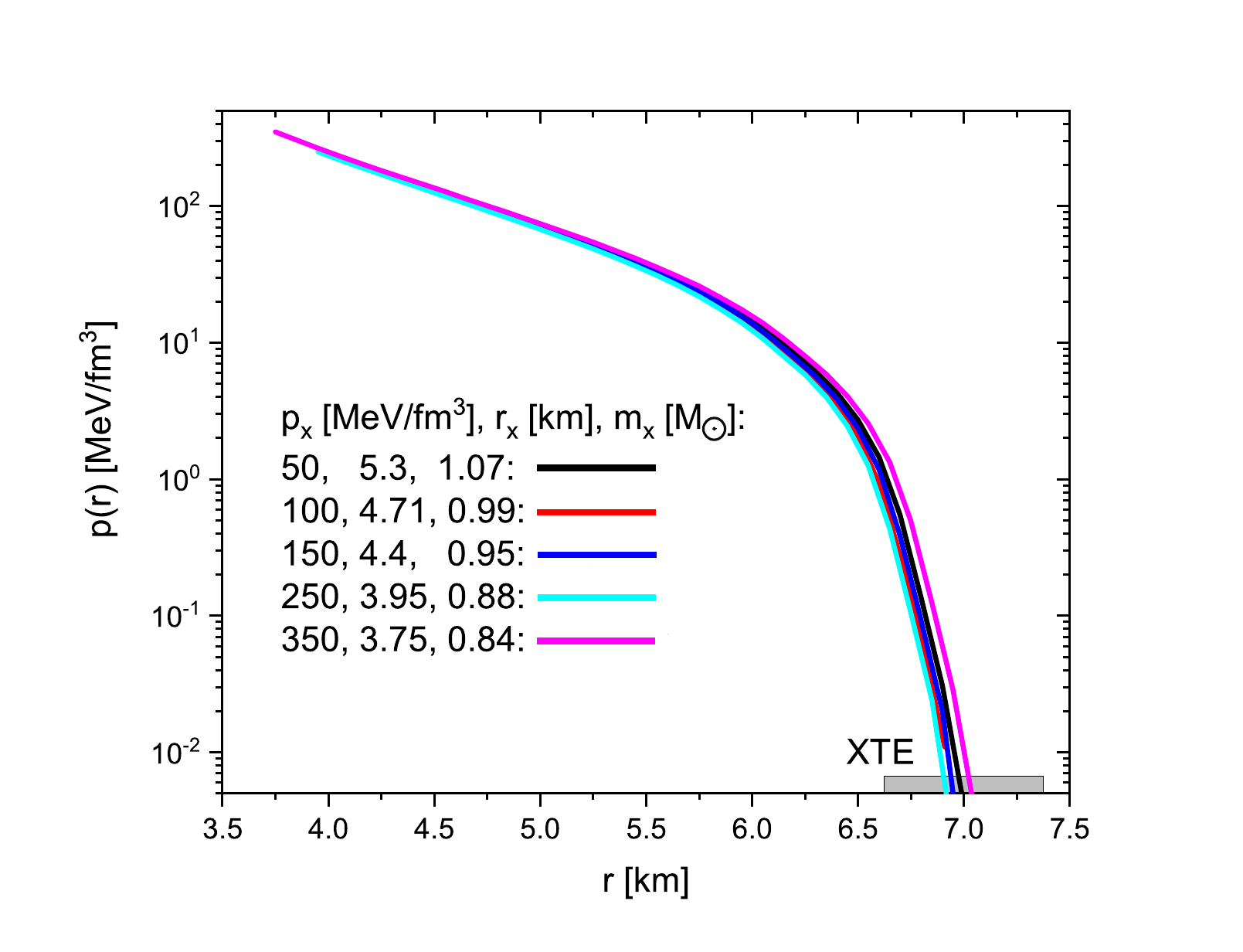} \hspace{-3mm}
	\includegraphics[width=0.49\columnwidth]{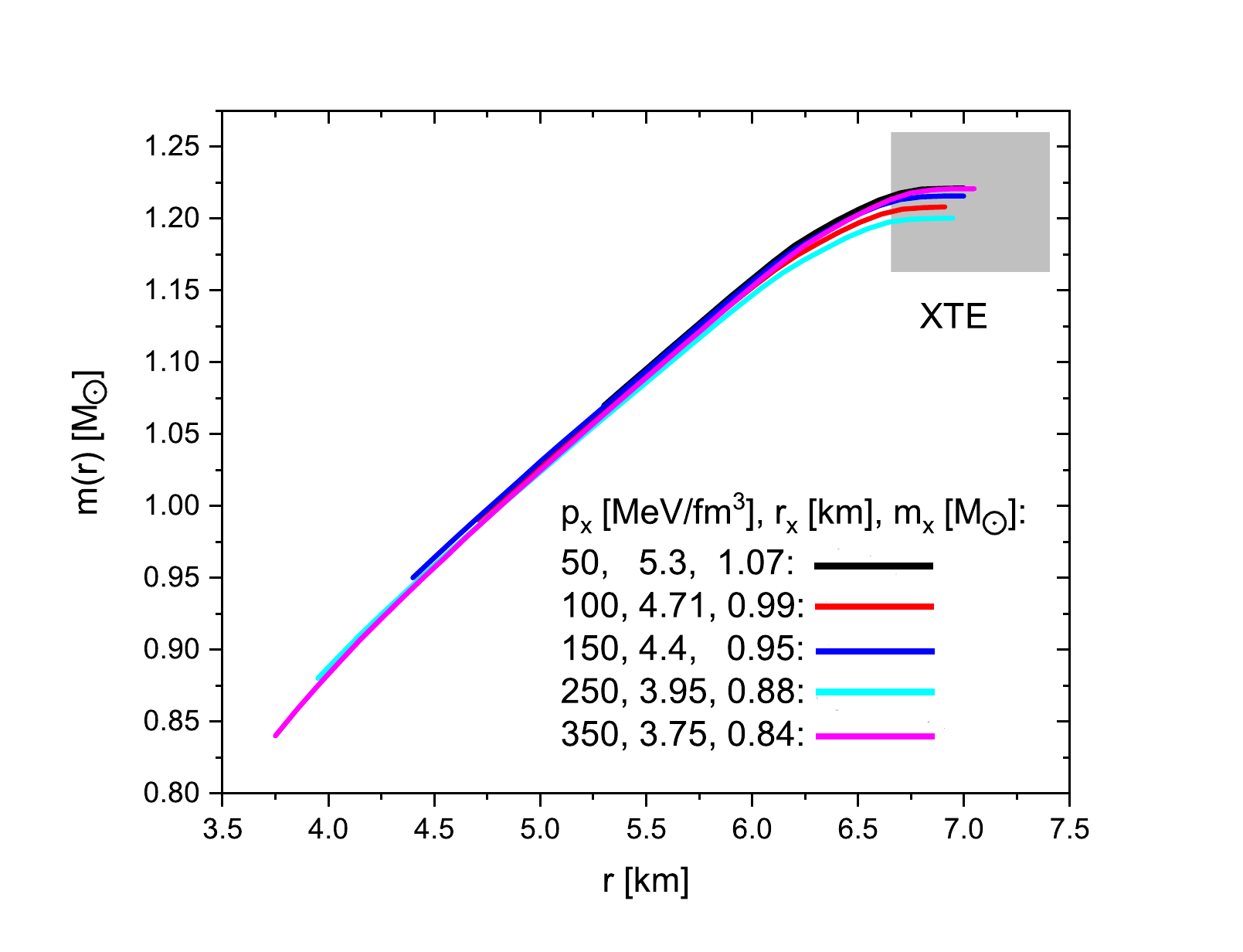} 
	\\[-6mm]
	\caption{{Profiles of} 
		pressure $p (r)$ (\textbf{left panel}) and mass function $m (r)$ (\textbf{right panel}) in the corona
		for various values of $p_x = 50, 100, 150, 250$ and $350~\mathrm{MeV/fm}^3$
		with values of $r_x$ and $m_x$ to satisfactorily match the current data of the XTE square
		(gray squares).
		The EoS is NY$\Delta$ as in Figures~\ref{fig:1} and \ref{fig:1b}.
		\label{fig:2} 
	}
\end{figure}

\begin{figure}[H]
		\centering
	\includegraphics[width=0.49\columnwidth]{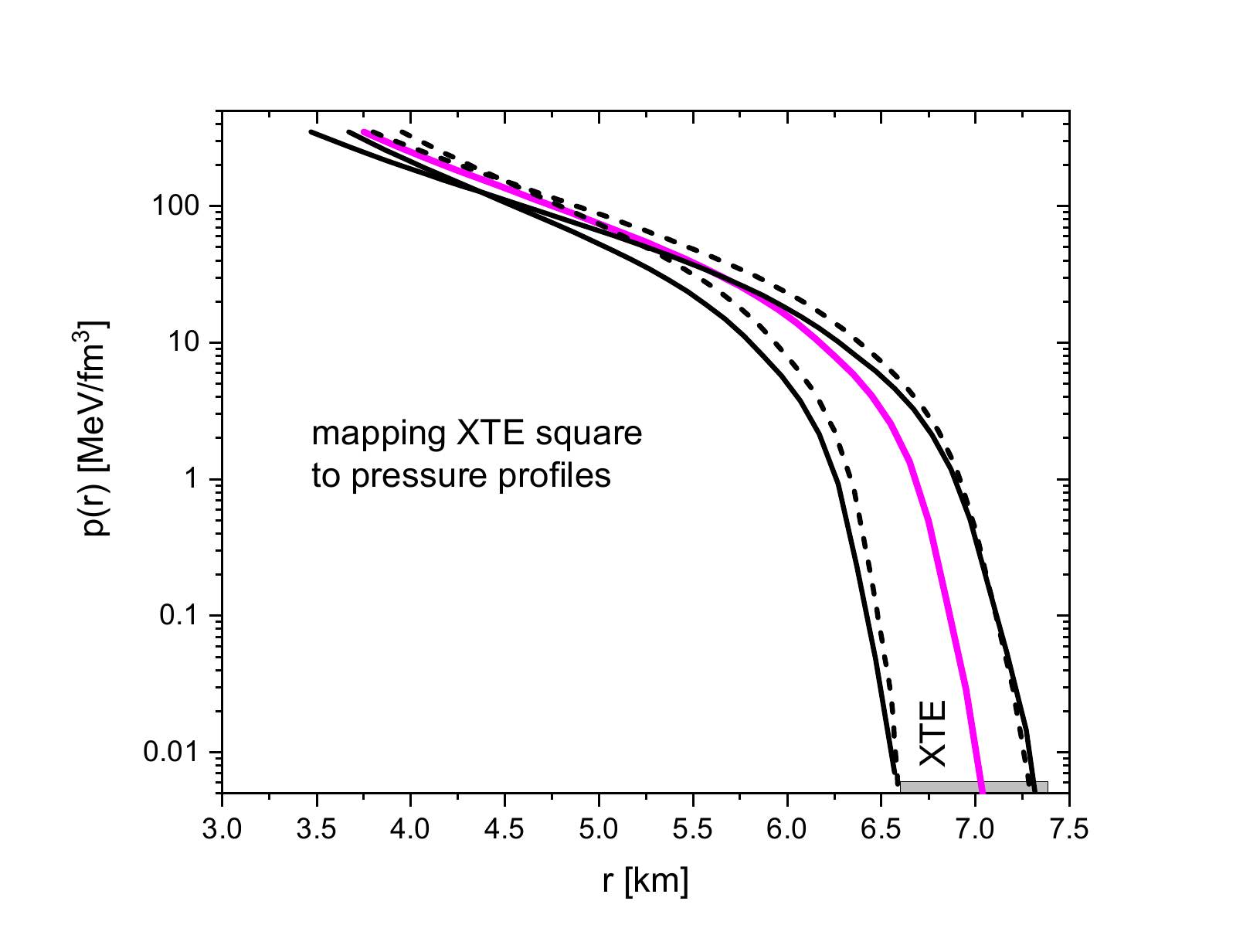} \hspace{-3mm}
	\includegraphics[width=0.49\columnwidth]{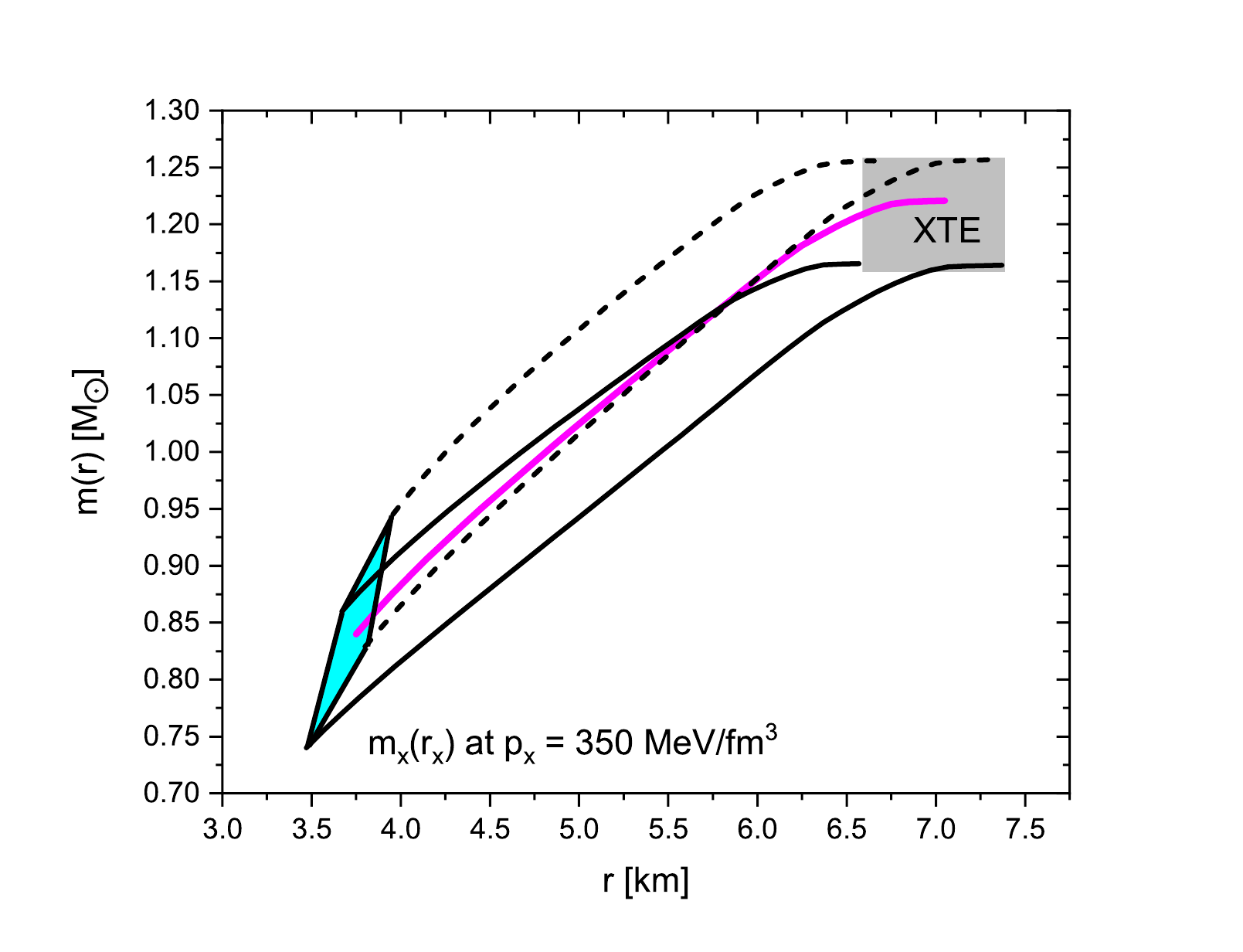}
	\caption{{Profiles of pressure $p (r)$ (\textbf{left panel}) and mass function $m (r)$ (\textbf{right panel}) in the corona
			for $p_x = 350~\mathrm{MeV/fm}^3$
			with values of $r_x$ and $m_x$ to satisfactorily match the corners of the XTE square {defined in Section \ref{masses}}.
			Dashed (solid) curves are for the maximum (minimum) masses.
			In the right panel, the~admissible area $m_x(r_x)\vert_{p_x = 350~\mathrm{MeV/fm}^3}$ is exposed
			(cyan parallelogram).
			For comparison, a~representative curve (fat magenta) is also shown for the respective quantities,
			approximately matching the center of the XTE square.
			The EoS is NY$\Delta$ as in Figures~\ref{fig:1}-\ref{fig:2}.
			\label{mapping}
	}}
	\end{figure}

The pressure and mass profiles {are} exhibited in Figure~2 in~\cite{Zollner:2024ufa}, also including 
the case of the HESS remnant J1731-347, both for $p_x = 50~\mathrm{MeV/fm}^3$.
Here, we display in Figure~\ref{fig:2} the profiles for $p_x = 50, 100, 150, 250$ and $350~\mathrm{MeV/fm}^3$
for XTE J1814-338 only. In~doing so, {by the shooting method}, we {select} the core parameters
$(p_x/(\mathrm{MeV/fm}^3), r_x/\mathrm{km}, m_x/M_\odot) = (50, 5.3, 1.07)$ (black curves),
$(100, 4.71, 0.99)$ (red curve),
$(150, 4.4, 0.95)$ (blue curve),\newline
$(250, 3.95, 0.88)$ (cyan curves) and
$350, 3.75, 0.84)$ (magenta curves), which {deliver} $(M, R)$ within the XTE square. 
Both the pressure profiles and the mass profiles are nearly on top of each other, suggesting the occurrence
of a master curve, {which could be obtained by inward integration}. However, NY$\Delta$ {is} limited to $p \le 373.58~\mathrm{MeV/fm}^3$,
i.e.,\ larger values of $p_x$ {are} not~accessible.
\noindent Analog master curves map the corners of the XTE square to $p(r)$ and $m(r)$;
	see \mbox{Figure~\ref{mapping}}. This accomplishes the mapping of the XTE square $\mapsto$ $(m_x, r_x)\vert_{p_x}$. %EE: Please check that intended meaning has been retained. 
	One admissible parallelogram is shown for $p_x = 350~\mathrm{MeV/fm}^3$ in the right panel.
	For each value of $p_x$, a~corresponding parallelogram can be constructed by combining the information
	depicted in both panels. Figure~\ref{fig:1b} above uses XTE ellipses and a few discrete values
	of $p_x$. The~mentioned approximations refer to the upper dashed and lower solid curves
	in Figure~\ref{mapping}(right).
\section{Considering {One-Fluid} Core~Examples}\label{app:B}
The CCD {was} useful for {isolating the} details of the corona.
What {remained} for further progress {was} to attempt an explanation of the core parameters $(p_x, r_x, m_x)$.
As pointed out in the previous subsection, one {has} to employ another EoS in the region of
$p~>~350~\mathrm{MeV/fm}^3$. We now {test} two examples for a core EoS:
(i)  an EoS related to the QCD trace anomaly (Section \ref{sec:DeltaEoS})
and (ii) a strong {first-order} phase transition (Section \ref{sec:FOPT}).
Both examples assume the applicability of the above {one-fluid} TOV~equations.
\subsection{Using Core Model with EoS Related to QCD Trace~Anomaly}\label{sec:DeltaEoS} 
\begin{figure}[H]
\centering
	\includegraphics[width=0.45\columnwidth]{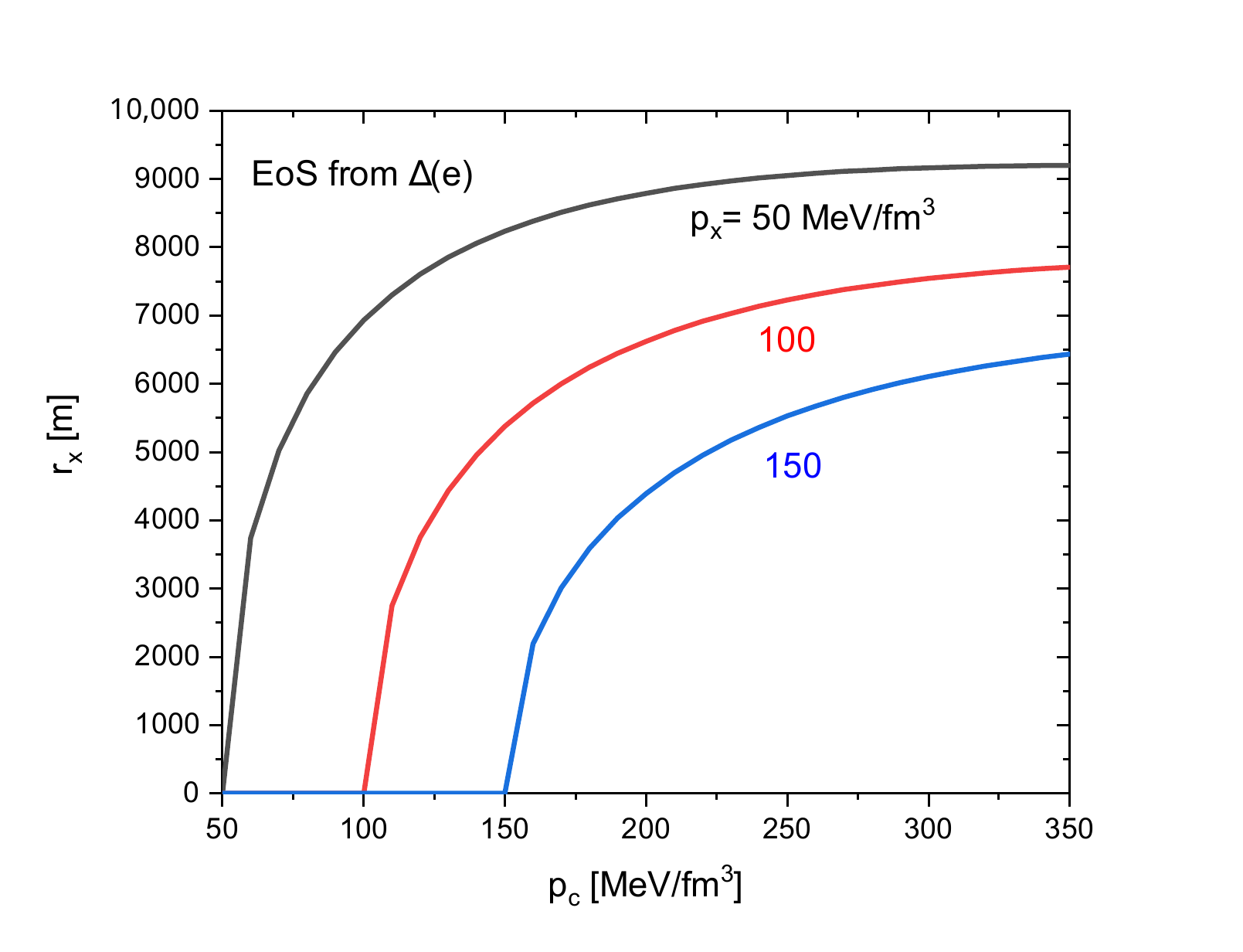}  \hspace{-5mm}
	\includegraphics[width=0.45\columnwidth]{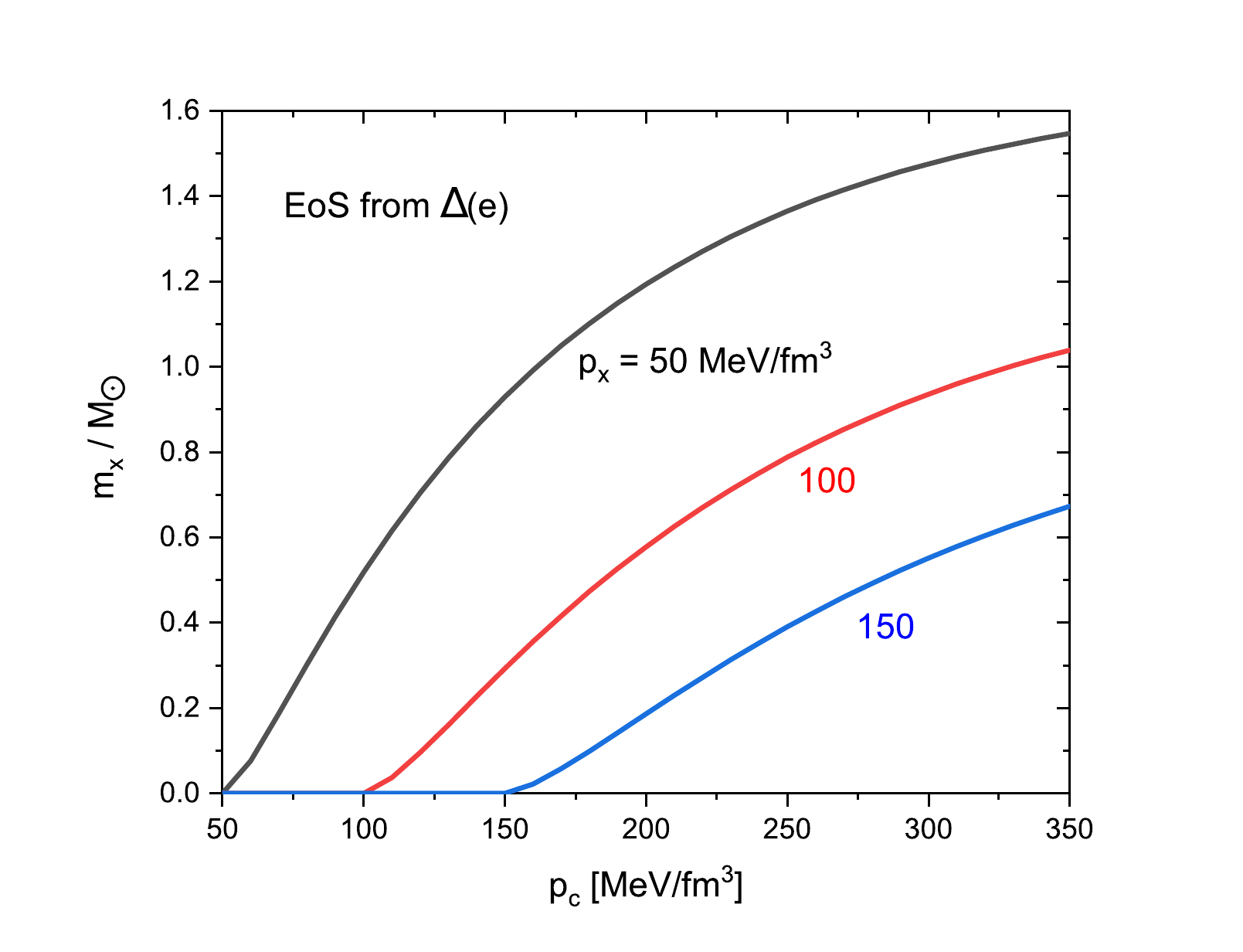} 
	\\[-6mm]
	\caption{{A}
		{one-fluid} core model
		using the EoS $p(e) = e [\frac13 - \Delta(e)]$, where $\Delta(e)$ is a fit (see Equation~(\ref{footnote.Delta}))
			of the results in~\cite{Marczenko:2023txe}
			based on a statistically determined EoS from {multi-messenger} data. 
			Core radii $r_x (p_c)$ (left panel) and masses $m_x (p_c)$ (right panel) 
			as a function of the central pressure $p_c$ for various 
			values of the pressure $p_x = 50, \, 100$ and $150~\mathrm{MeV/fm}^3$ (from top to bottom)
			at the core surface, where $p(r_x) = p_x$
			}.
		\label{fig.Delta} 
	\end{figure}

To present an explicit example of a {one-fluid} core, we {adapt} the EoS related to the QCD trace anomaly $\Delta$, i.e.,\
$p(e) = e [\frac13 - \Delta(e)]$ and squared sound velocity
$v_s^2 = \frac13 - \Delta - e \frac{\partial \Delta}{\partial e}$.
In~\cite{Marczenko:2023txe}, $\Delta (e)$ is statistically determined
with constraints from {multi-messenger} neutron star data {with respect to} hints of approaching conformality 
in the cores.
$\Delta(\eta)$ with \mbox{$\eta := \ln (e/ 150~\mathrm{MeV/fm}^3)$},
originally proposed in~\cite{Fujimoto:2022ohj}, {agreed} -- with a tiny correction of one parameter -- with $\Delta^{\mathrm{NY}\Delta} (\eta > 0.9)$ from~\cite{Li:2019fqe}; see Figure~3-left in~\cite{Zollner:2023myk}.
The plots of $\Delta (e)$ and $v_s^2 (e)$ in~\cite{Marczenko:2023txe} (see Figure~4 there) can be parameterized by
\begin{align} \label{footnote.Delta}
	\Delta = 0.33 \left[1-\frac{A}{1+\exp\{-\kappa (\eta - \eta_c) \} } \right]
	+ G_G \exp \left\{- \frac{(\eta - \eta_G)^2}{2 \sigma_G} \right\}
\end{align} 
with $\eta := \ln (e / \hat e_0)$,
$\hat e_0 = 0.12~\mathrm{GeV/fm}^3$,
$A = 1.52$,
$\kappa = 3.582856$,
$\eta_c = 1.357143$,
\mbox{$G_G = 0.2926$}, 
$\eta_G = 4.045714$ and
$\sigma_G = 2$.\\
The {one-fluid} TOV equations 
{are} integrated from $r = 0$,
where $p = p_c$ and $m = 0$,
to $r_x$, where $p (r_x) = p_x$ and $m_x = m (r_x)$.
The emphasis here {is placed} on the {non-trivial} dependence of both
$\Delta (e)$ and  $v_s^2 (e)$, which {encode} the EoS. %EE: Please check that intended meaning has been retained. 
The corresponding core radii and masses 
as functions of the central pressure $p_c$ {are} {shown} in Figure~\ref{fig.Delta}.
Both core radii and core masses {increased} with central pressure. 
(For comparison, somewhat smaller values of $r_x(p_c)$ and $m_x(p_c)$
would {have been} obtained when using NY$\Delta$ as a core EoS.)
Note that, for~a {one-fluid} core with a given EoS and $p_x$, the~values of $r_x$ and $m_x$ become
correlated due to the $p_c$ dependence, and~the usual stability criteria apply\footnote{A specific model of a core of asymmetric DM surrounded by an SM envelope is presented in
	Figure~9-right in~\cite{Gresham:2018rqo}. (We thank J.~Schaffner-Bielich for bringing this reference
	to our attention.)
	The explicitly given EoSs of core and corona allow for stability analyses and tidal deformability evaluations.}.
	\begin{figure}[H]
		\centering
		\includegraphics[width=0.6\columnwidth]{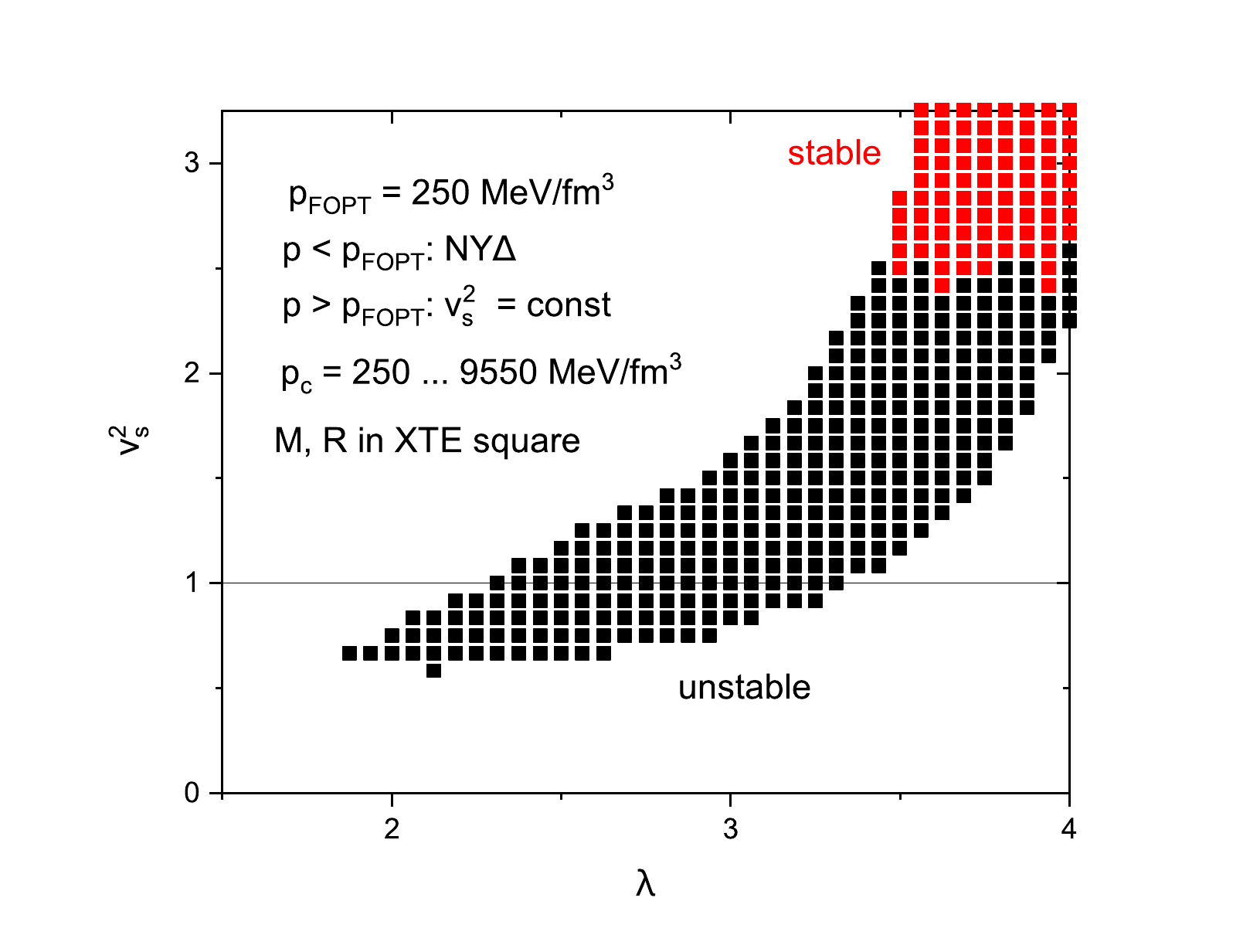} 
		\\[-6mm]
		\caption{
			{Values} of $v_s^2$ and $\lambda$, where the combination of NY$\Delta$ with an FOPT model
			delivers $M(p_c)$ and $R(p_c)$ within the XTE square.
			Only in the region of red squares are the XTE compatible configurations stable, i.e.,\
			$M(R)$ increases with increasing $p_c$. %EE: Please check that intended meaning has been retained. 
			In this region, however, the~{high-density} EoS violates causality since $v_s^2 > 1$.
			\label{fig.vs2-lambda} 
		}
	\end{figure}
\noindent This example of a {one-fluid} core model {did} not deliver consistent values 
of $r_x(p_c; p_x)$ and $m_x(p_c; p_x)$ at $p_x = 50, 100$ and $150~\mathrm{MeV/fm}^3$,
which {were} needed -- in combination with the NY$\Delta$ corona model -- to match the XTE J1814-338 data~\cite{Kini:2024ggu};
see Figure~\ref{fig.vs2-lambda}.
{For XTE J1814-338,} a more compact core {is required}, which could {be} accomplished using a DM admixture
{by} a strong {first-order} phase transition (FOPT).
\subsection{{First-Order} Phase~Transition}\label{sec:FOPT}
In this subsection, we {consider} an example of an FOPT.
That is, the~NY$\Delta$ EoS is continued at $p > p_\mathrm{FOPT}$
by a constant sound velocity model EoS.
Following~\cite{Laskos-Patkos:2024fdp}, we continue the NY$\Delta$ EoS 
at $p > p_\mathrm{FOPT} = 250~\mathrm{MeV/fm}^3$
with
\begin{align}
	e(p) = \lambda e^{\mathrm{NY}\Delta}(p_\mathrm{FOPT}) + v_s^{-2} (p - p_\mathrm{FOPT}) .
\end{align}
This choice of  $p_\mathrm{FOPT}$ {left the stable NY$\Delta$ branch unaffected}
(see the {bold} solid black curve in Figure~\ref{fig:1}) with
$R \approx 12~\mathrm{km}$ and masses up to $2.02~M_\odot$, thus being (marginally) consistent with
many gravitational waves, {multi-messengers} and NICER observations. %EE: Please check that intended meaning has been retained. 
A sufficiently strong FOPT with an energy density jump parameterized by $\lambda$ at $p_\mathrm{FOPT}$
and constant sound velocity squared $v_s^2$ at $p > p_\mathrm{FOPT}$
initially {bent} the $M(R)$ curve down. 
For running $p_c$ up to extraordinarily large values in the order of
$10~\mathrm{GeV/fm}^3$, one {could} catch the XTE square (see Figure~\ref{fig:1}).
Figure~\ref{fig.vs2-lambda} {shows} the region $v_s^2$ vs.\ $\lambda$, where $M$ and $R$ fall in the XTE square.
In contrast to~\cite{Laskos-Patkos:2024fdp}, our EoS combination {did} not allow for stable compact (neutron) stars 
with causal {high-density} EoS. \\
Despite this failure, {attention may still be directed toward} {core-only} properties and use the above EoS construction.
Keeping $p_\mathrm{FOPT} = 250~\mathrm{MeV/fm}^3$ and selecting 
$p_c = 5~\mathrm{GeV/fm}^3$ together with $\lambda = 2.9$ and $v_s^2 = 1$,
we {determined} the corresponding core, i.e.,\ radii $r_x$ determined by $p(r_x) = p_x$. 
Interestingly, for~$p_x = 50, 100, 150$ and $250~\mathrm{MeV/fm}^3$, the~{core-mass-core-radius relationship}
{was approximately} $m_x \approx (0.33 + 0.14 \frac{r_x}{\mathrm{km}}) M_\odot$, which {matched} the $p_x$ dependence of
the XTE ellipse centers displayed in Figure~\ref{fig:1b}. That is, such a model EoS and such a $p_c$ choice
would {have served} as a specific core model. The~special case of $p_x = 250~\mathrm{MeV/fm}^3$ is a model
with a core as a {high-density} EoS and a corona as nuclear (NY$\Delta$) matter, both joined by an FOPT.\\
Both examples {demonstrate} that a {one-fluid} SM matter core {is} unlikely to explain the required core
parameters for XTE J1814-338. In~Appendix \ref{app:C}, we {consider} the special case of a ``trivial core'',
which means that a distinguished core {is} not~needed. 
\section{Summary} \label{sect:summary}
The CCD {relied} on the agnostic assumption that the EoS of compact (neutron) star matter 
(i) {was reliably known} up to energy density $e_x$ and pressure $p_x$ and 
(ii) SM matter {occupied} the star as the only component at radii $r > r_x$. 
The baseline for static, spherically symmetric configurations {was then} provided by the TOV equations,
which {were} integrated, for~$r \in [r_x, R]$, to~find the circumferential radius $R$ (where $p(R) = 0$) and the
gravitational mass $M = m(R)$. 
This {accomplished} the mapping $(M, R) \mapsto (m_x, r_x)\vert_{p_x}$.  
We {called} the region $r \in [r_x, R]$ ``corona'', 
but ``crust'' or ``mantle'' or ``envelope'' or ``shell'' or ``halo''
are also suitable synonyms. The~region $r \in [0, r_x]$ {was} the ``core'', parameterized by 
the included mass $m_x$. The~core {had to} support the corona pressure at the interface, i.e.,\ $p(r_x^-) = p(r_x^+)$,
assuming either pure SM matter or a DM component  {feebly interacted} with the SM matter component. 
The core {could have contained} any material compatible with the symmetry requirements.
In particular, it could {have been} modeled by {multi-component} fluids with SM matter plus DM
or {any other form of matter beyond} the SM. 
Alternatively, an~FOPT {could have been} accommodated in the SM {matter-only one-fluid} core.
Then, $p_x = p_\mathrm{FOPT}$ {would have been} appropriate; see Section \ref{sec:FOPT}.\\
The currently available {mass-radius} data of XTE J1814-338 
{point} to an averaged core mass density
$\langle \rho \rangle_x := 3 m_x /(4 \pi r_x^3) \approx 3 \cdot 10^{15}~\mathrm{g/cm}^3$
and seem to belong to another class of very compact (neutron) stars.\\
The tidal deformability and stability properties {remained} challenging issues. 
Improved data {would} provide further constraints and pave the way {toward} explicating the core properties
{with respect to} the options of an FOPT and/or DM~admixtures.\\
In some sense, our results {supported} the model of a {two-family} approach
based on two distinct classes of EoSs~\cite{DiClemente:2021dmz}.\\
While the accomplished decomposition {seemed} to leave the determination of the advocated core by explicitly
accommodating either an SM matter EoS or an SM + DM mixture with separate EoSs to reproduce the
triple $(r_x, m_x; p_x)$, our CCD construction {is} not yet universal since it {depends} on the actually employed
corona EoS. In~{follow-up} work, one {has to} test the robustness of the deduced values $(r_x, m_x; p_x)$
by using other methods than the NY$\Delta$ EoS.  
In particular, we {envisage}
replacing the deployed EoS NY$\Delta$ with more refined models, which also {catch} high masses
up to $2.35~M_\odot$. 
\subsection*{Acknowledgement}
One of the authors (B.K.) acknowledges continuous discussions with
J.~Schaffner-Bielich and K.~Redlich for encouragement to deal with the current topic.
We thank A.~Bauswein and W.~Weise for useful~conversations. In addition, we thank Ella Jannasch for her virtuoso handling of all programs during manuscript processing.

\begin{appendices}
	\section{What If No Core Is Required?}\label{app:C}
\begin{figure}[H]
	\includegraphics[width=0.49\columnwidth]{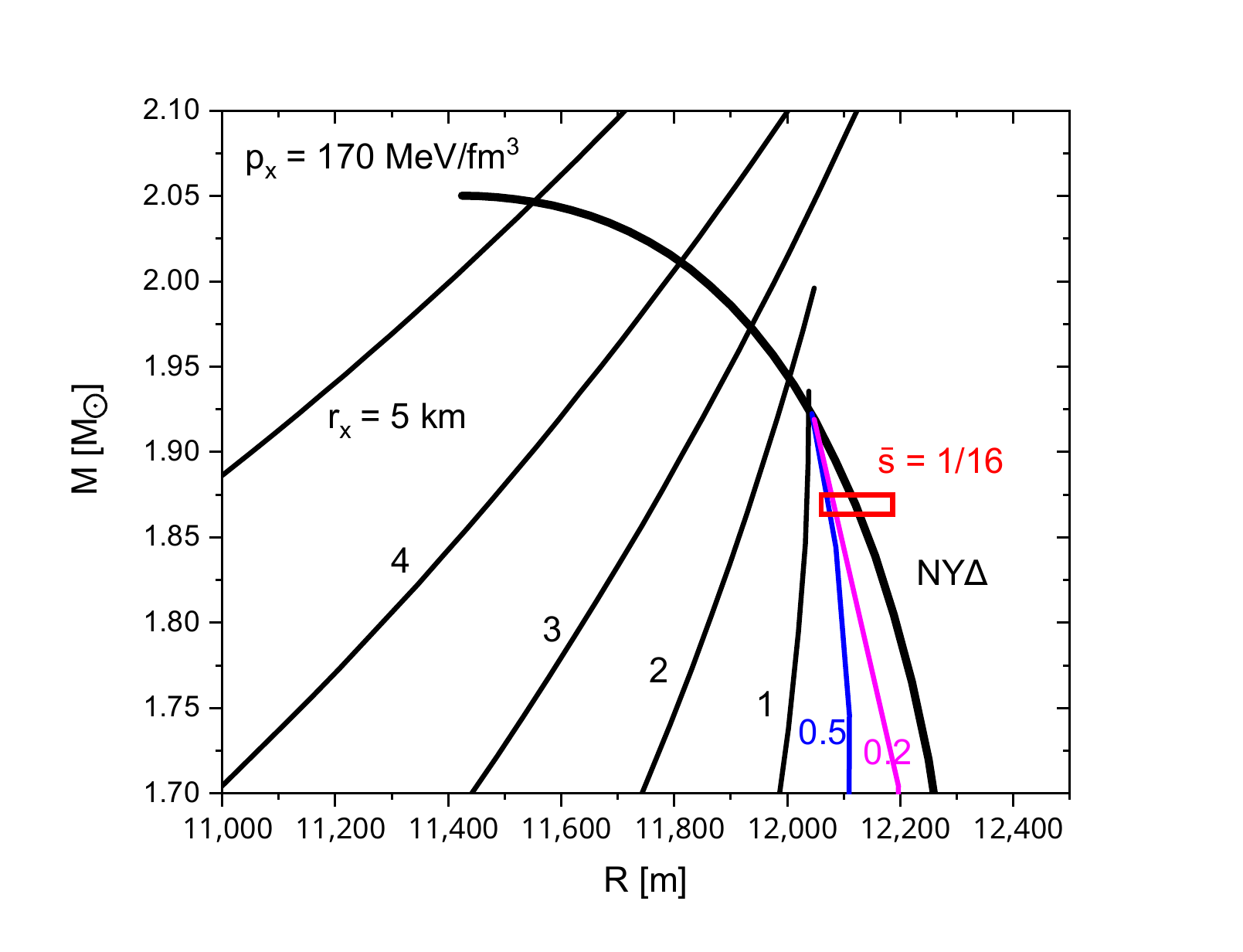} 
	\includegraphics[width=0.49\columnwidth]{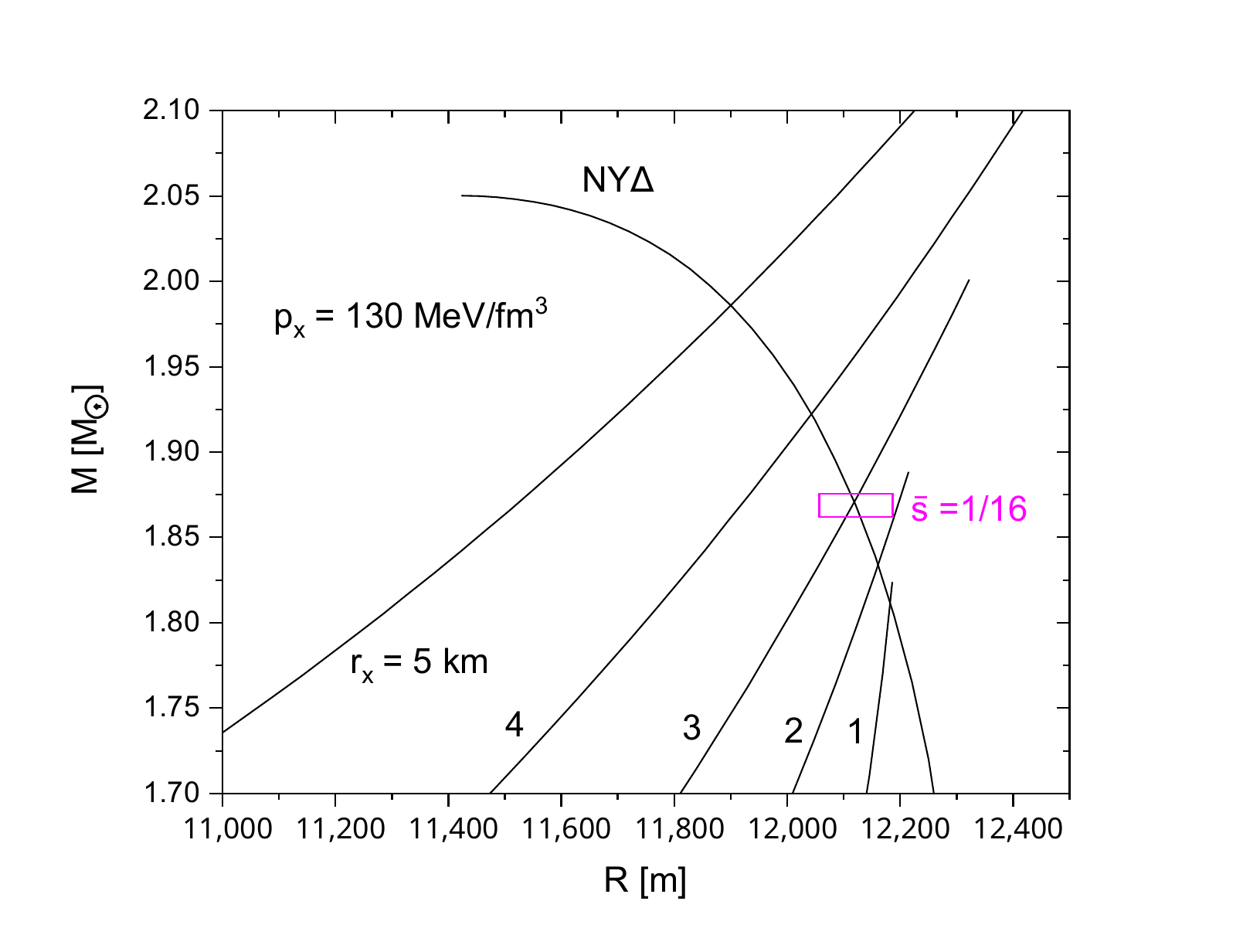}  
	\\[-6mm]
	\caption{{{Mass-radius}
			relationships} for CCD with $p_x = 170$ (\textbf{left panel})  
		and $130~\mathrm{MeV/fm}^3$ (\textbf{right panel}).
		The right/upper end points are for $m_x = 0$;
		$m_x$ increases when going to left/down on the curves with $r_x = const$.
		The size of the assumed credibility region (red rectangle)
		is steered by $\bar s = 1/16$.
		The fat solid curve is for NY$\Delta$ with varying $p_c$.
		\label{fig.S2c} 
	}
\end{figure}

\begin{figure}[H]
	\centering
	\includegraphics[width=0.49\columnwidth]{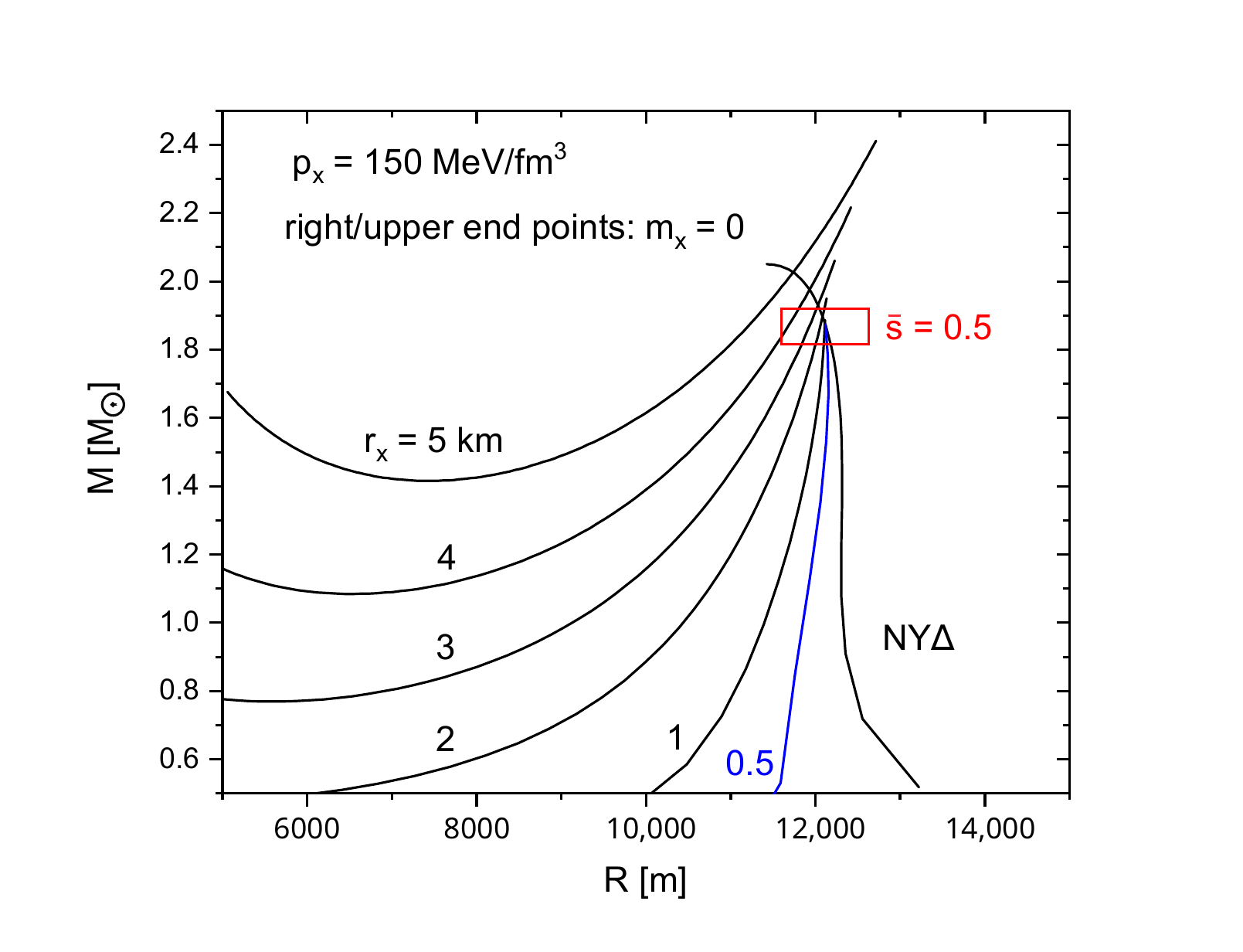} 
	\includegraphics[width=0.49\columnwidth]{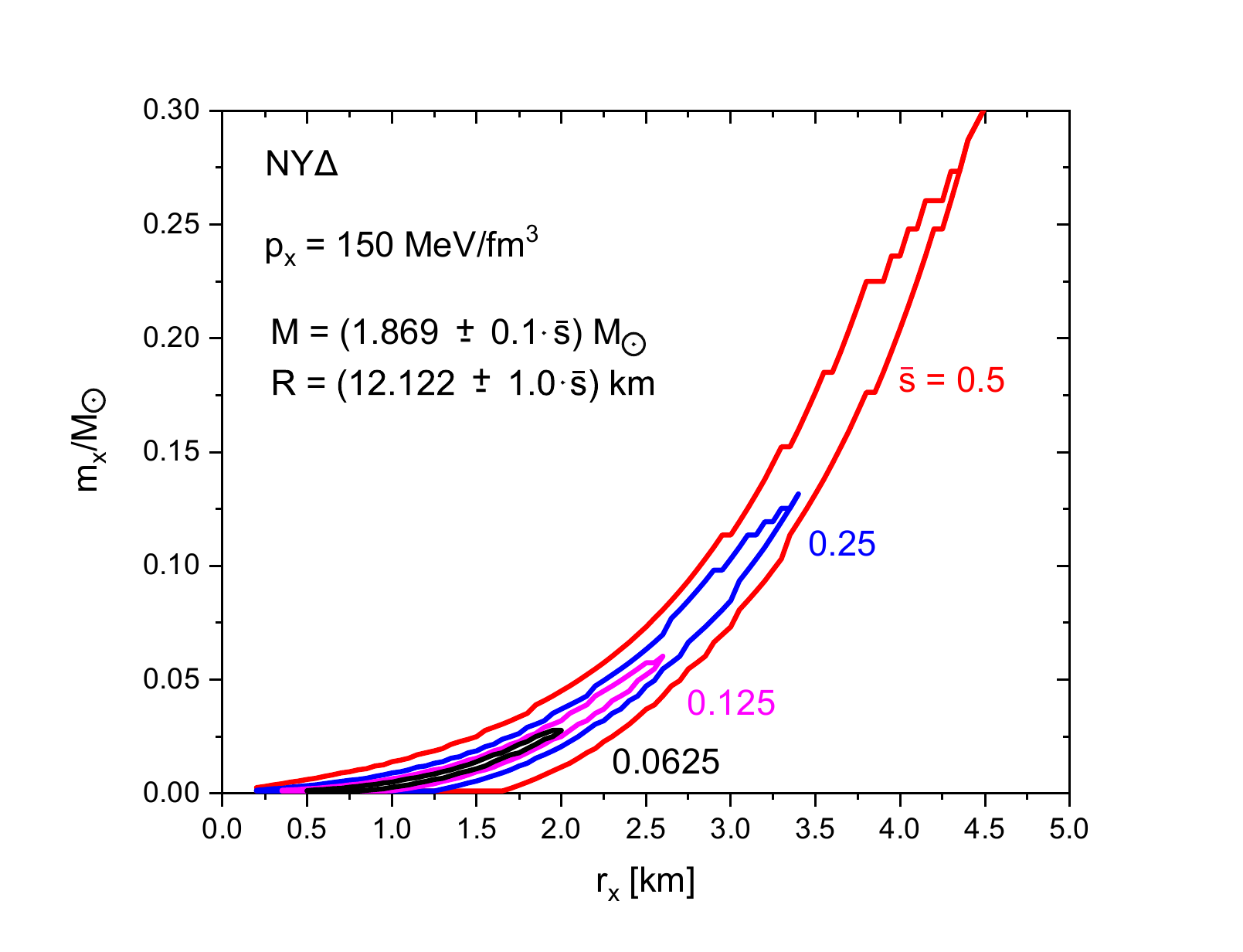}
	\\[-6mm]
	\caption{(\textbf{{left panel}
		})
		As in Figure~\ref{fig.S2c} but for $p_x = 150~\mathrm{MeV/fm}^3$ and $\bar s = 1/2$.
		(\textbf{{right panel}})
		{Areas of $m_x$ and $r_x$ delivering, via {one-fluid} TOV equations with NY$\Delta$ 
			and for $p_x = 150~\mathrm{MeV/fm}^3$,
			compact (neutron) stars within the NY$\Delta$ squares
			for $\bar s = 0.5$ (red), 0.25 (blue), 0.125 (magenta) and  0.0625~(black).}
		\label{fig.S2} 
	}
\end{figure}

To see how the CCD would {work} if no core {were} required, we {defined} the ``NY$\Delta$ square'' by
$M \in (1.869 \pm 0.1 \bar s)~M_\odot$ and $R \in (12.122 \pm 1 \bar s)~\mathrm{km}$.
In the center of this square, $\bar s = 0$, {was} the pure NY$\Delta$ configuration with $p_c = 150~\mathrm{MeV/fm}^3$,
depicted in Figure~\ref{fig:1} by the~asterisk. \\
First, we {supplemented} the information of Figure~\ref{fig:1} by exhibiting in Figure~\ref{fig.S2c}
{the} CCD for $p_x =170$ (left panel) and $130~\mathrm{MeV/fm}^3$ (right panel). 
Some curves with $r_x = const$ {crossed} the NY$\Delta$ square for $\bar s = 1/16$.
The overall pattern is analogous to  Figure~\ref{fig:1} but enriched by curves with $r_x = const$
for $p_x = 150~\mathrm{MeV/fm}^3$ displayed in Figure~\ref{fig.S2}(left) together with the
NY$\Delta$ square for $\bar s = 1/2$.
\noindent Keeping $p_x = 150~\mathrm{MeV/fm}^3$ and requiring  $(M, R)$ in the NY$\Delta$ square
for \linebreak $\bar s = 0.5, \ldots, 0.0625$, one {finds} the admissible regions in the $m_x$ vs. $r_x$ plane 
exhibited in Figure~\ref{fig.S2}(right). 
The upper boundaries {were} determined by $M_{\min}$ (the lower mass limit of the NY$\Delta$ square)
at lower values of $(m_x, r_x)$, while,
at larger values of  $(m_x, r_x)$, $R_{\min}$ (the lower radius limit of the NY$\Delta$ square)
{was} the steering quantity. The~turn was about at $r_x \approx 3.7$~km
for $\bar s = 0.5$. The~lower boundary {was} steered by $M_{\max}$ (the upper mass limit of the NY$\Delta$ square).
This behavior {could be} inferred by the
left panel upon inspecting the entry and exit points of the curves $r_x = const$ entering and leaving the respective
NY$\Delta$ square. Note that $R_{\max}$ (the upper radius limit of the NY$\Delta$ square)
{was} not relevant~here.\\
{Mimicking} the increasing accuracy of $(M, R)$ data 
by reducing the parameter $\bar s$, the~admissible areas {shrank}.
$\bar s = 1$ roughly {corresponded} to a 10~\% accuracy, both in $M$ and $R$.
Ideally, $\bar s \to 0$ {would have resulted} in $m_x \to 0$ and  $r_x \to 0$ since, for~$p_c = p_x$, no core is required.
{Supposing} a highly reliable EoS {was} at our disposal, one {could have deduced} in this spirit whether a {finite-sized} core
{could have been} accommodated in certain $(M, R)$ data. Such a core {might have been} considered as signaling DM (admixture). 
For the present set-up, however, the~approach of $(m_x, r_x) \to 0$ with increasing values of $\bar s$
clearly {told} us that a distinguished core {was} not~needed.
\begin{figure}[H]
	\centering
	\includegraphics[width=0.49\columnwidth]{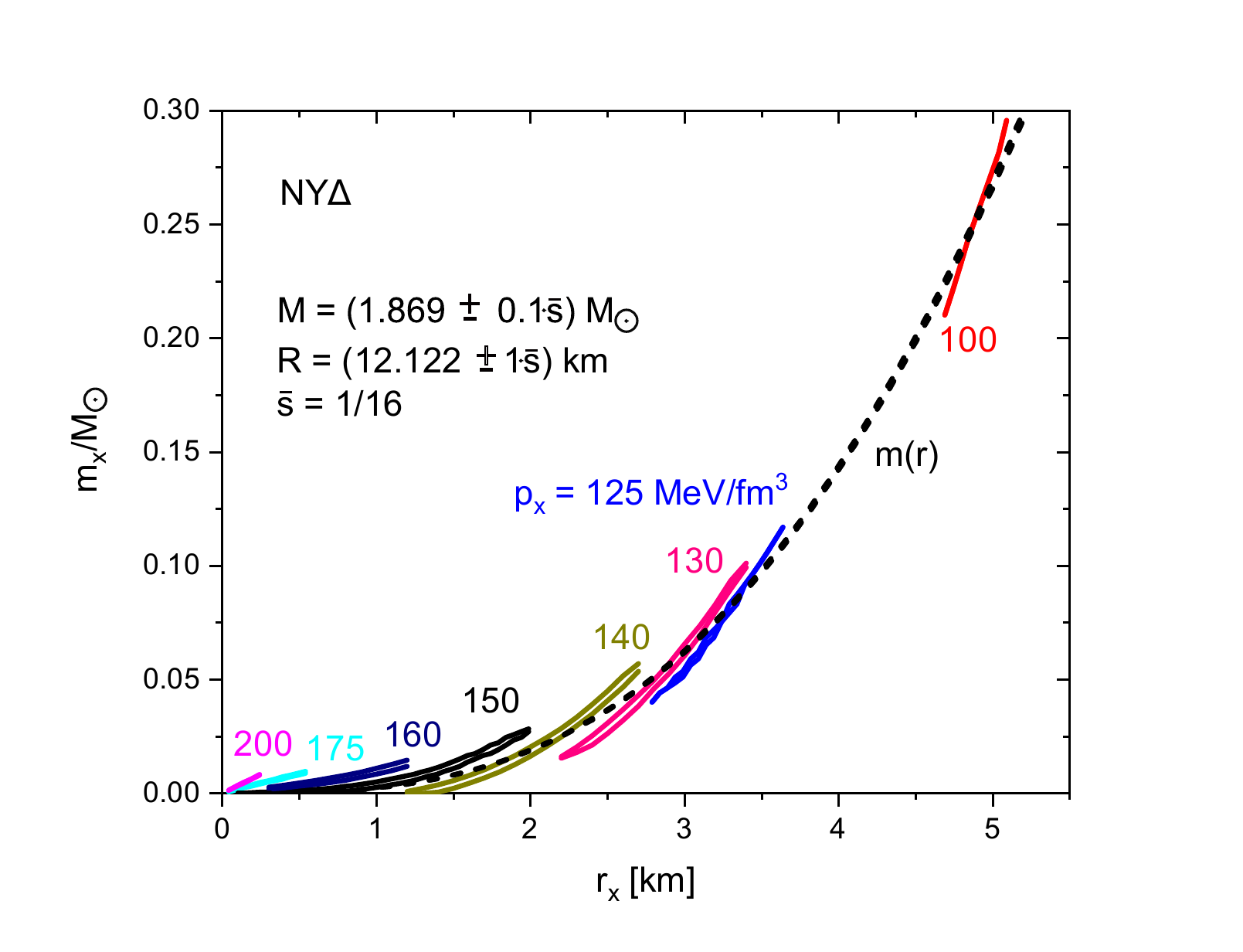}   
	\includegraphics[width=0.49\columnwidth]{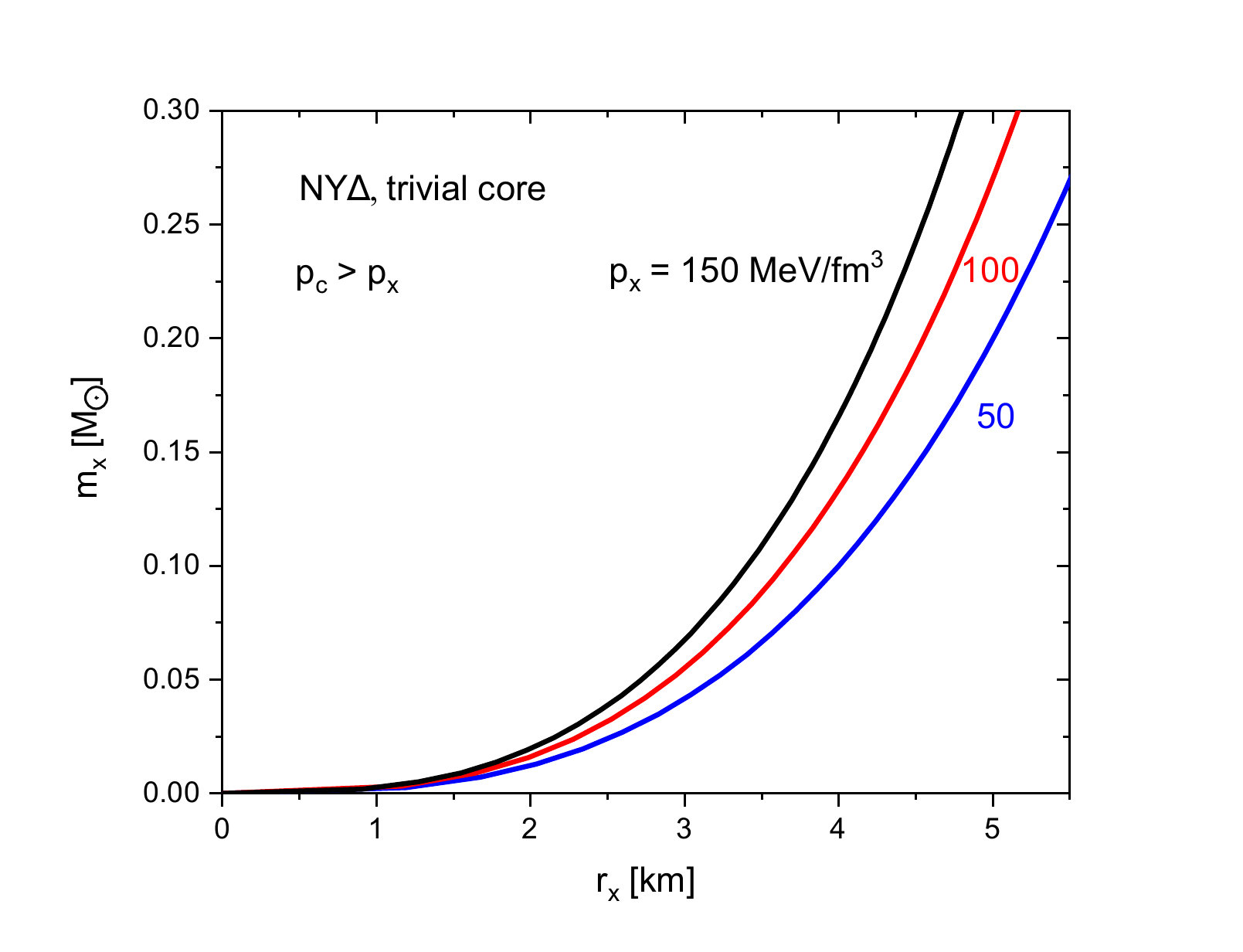} 
	\\[-6mm]
	\caption{(\textbf{{left panel}})
		The narrow areas of $m_x$ and $r_x$ delivering, via {one-fluid} TOV equations with NY$\Delta$ 
		and for various values $p_x = 100, \ldots, 200~\mathrm{MeV/fm}^3$,
		compact (neutron) stars with $(M, R)$ within the NY$\Delta$ square with $\bar s = 1/16$.
		The pattern can be inferred from Figures~\ref{fig.S2c}~{(left)} 
		and \ref{fig.S2}~(left).
		The lower (upper) dashed curve is for $m(r)$ with NY$\Delta$ and $p_c = 147.9 ~(152.2)~\mathrm{MeV/fm}^3$.
		(In the displayed scale, both curves are nearly on top of each other.)
		(\textbf{{right panel}})
		{Core-mass vs.\ core-radius relationships} $m_x(r_x)$ for
		``trivial cores'' with NY$\Delta$ resulting from {one-fluid} TOV equations integrated from
		$p_c$ to $p_x$, where $r_x (p_c)$ and $m_x(p_c)$ are found, for~$p_x = 50$ (blue curve),
		$100$ (red curve) and $150~\mathrm{MeV/fm}^3$ (black curve).
		\label{fig.S2a} 
	}
\end{figure}
\noindent In Figure~\ref{fig.S2a}~(left), the~needed values of $m_x (r_x)$ {are} exhibited for various values of~$p_x$.
For $\bar s = 1/16$, the~credibility regions {shrink} and {move} towards $m_x \to 0$ and $r_x \to 0$ for increasing
values of $p_x$. In~fact, the~CCD analysis of the corona signals that no core is needed for the trivial case in
which $(M,R)$ are given by NY$\Delta$ alone. 
The dashed curves are the mass profiles $m(r)$ for $p_c = 147.9$ and 
$152.2~\mathrm{MeV/fm}^3$. These limiting central pressures ensure that the pure NY$\Delta$
configurations (no core) {fall} within the NY$\Delta$ square. Interestingly, the~curves 
$m(r)\vert_{p_c \approx 150~\mathrm{MeV/fm}^3}$ {neatly intersect with} the admissible $m_x(r_x)\vert_{p_x}$ regions.\\
Note the striking difference to Figure~\ref{fig:1b}.
In the present case, the~credibility region uncovers $(M, R)$ provided by NY$\Delta$ with
$p_c = 150~\mathrm{MeV/fm}^3$, and~$m_x \to 0$, $r_x \to 0$ is conceivable,
meaning that for large values of $p_x$ no core is required and NY$\Delta$ applies {throughout}.
This is further exposed by stressing that 
$m_x = \frac{4 \pi}{3}A\frac{\bar \rho}{\mathrm{g/cm^3}} (\frac{r_x}{\mathrm{km}})^3 M_\odot$, 
\mbox{$A = 5 \cdot 10^{-17}$}, describes the trend very well for $\bar \rho = 10^{15}~\mathrm{g/cm}^3$. \\
While the left panel in Figure~\ref{fig.S2a} is purely related to the corona in CCD with a given $p_x$ at a given 
$(M, R)$ credibility region
(compared, however, to~the mass profile without reference to the core or the corona), the~right panel exhibits the
trivial core constructed by TOV integration from running $p_c$ to $p_x$ for various values of~$p_x$.
The latter construction also delivers the relation $m_x (r_x)$. Our {emphasis} here is again that
vanishing core mass and vanishing core radius are conceivable in a {straightforward} manner, thus completing our discussion
of a trivial (i.e.,\ unnecessary) core. 
In both cases, the~core compactness goes to zero for decreasing values of $r_x$, in~stark contrast to
Figure~\ref{fig:1b}, where the compactness~increases.\\
The need to allow for a noticeable core to accomplish given/measured pairs of $(M, R)$
in some credibility region can mean that
the EoS at high pressure fails, leaving many options unsettled, 
such as a failure of pure SM EoSs in {one-fluid} TOV equations or
DM (admixture) (requiring a {multi-fluid} TOV approach with various components or a pure DM core, where
{one-fluid} TOV equations with DM matter EoSs are appropriate).
In addition, our CCD relies on the assumption of an SM corona, which may be not applicable.
One should keep in mind, furthermore, that rotational effects and strong magnetic fields increase the complexity 
of possible scenarios.
\section{Adler's Simulated Horizon and Collins' Spiral}\label{AppB}
CCD is useful for other purposes too. For instance, Adler~\cite{Adler:2025jer,Adler:2023hxw}
considered black hole mimickers by studying a corona with a scale-free EoS
in the context of dynamical grava-stars~\cite{Adler:2022fqu}.
Zurek and Page~\cite{Zurek:1984zz} envisaged a black hole in/as the core.
The self-similar corona EoS $p = v_s^2 e$ allows to write the TOV equations as autonomous system
(cf.~\cite{Collins:1985}):
\begin{align}
	\frac{\mathrm{d} \alpha}{\mathrm{d} t} &= 3 \delta - \alpha, \label{aTOV1}\\
	\frac{\mathrm{d} \delta}{\mathrm{d} t} &= 2 \delta - 4 \delta \frac{\alpha + \delta}{1 - 2 \alpha}, \label{aTOV2} 
\end{align} 
where $\alpha := m/r$ and $\delta := 4 \pi r^2 p$ in geometric units, and we focus on constant
sound velocity squared $v_s^{-2} = 3$.
The radial coordinate $r$ is related to the quantity $t$
via $t - t_0 = \log(r/r_0)$ from $\mathrm{d} t = \mathrm{d} r /r$, where $t_0$ is an arbitrary constant,
and $r_0$ sets a length scale.
In the spirit of CCD, the core radius $r_x$ can be identified with $r_0$, and initial values
of $\alpha_0 = \alpha (t = t_0)$ and $\delta_0 = \delta (t = t_0)$ are determined by $m_x$ and $p_x$\footnote{
	Taking the example of $(r_x, m_x, p_x) = (3.75~\mathrm{km}, 0.84~M_\odot, 350~\mathrm{MeV/fm}^3)$
	one gets $\alpha_0 = 0.3322$ and $\delta_0 = 0.081789$.
	Note, however, that these core parameters refer to our CCD explanation of the XTE J1814-338 data
	with EoS NY$\Delta$ which is quite different to a scale-free EoS. 
}.
The advantage is the invariance against horizontal shifts encoded in $t_0$ which is chosen as zero.

Solving these autonomous equations accomplishes a systematic approach to Adler's
``simulated horizon''~\cite{Adler:2025jer,Adler:2023hxw}.
That is, for suitable initial conditions, the quantity $D := 1 - 2 \alpha$ displays a sharp kink
at a certain value of $t$, $t_{\mathrm{kink}}$,
and in some region $t > t_{\mathrm{kink}}$, the function $D \approx 1 - \exp (t_{\mathrm{kink}} - t)$ is nearly
indistinguishable from the exterior Schwarzschild solution
with $D = (1 -r_{\mathrm{kink}} / r)$~\cite{Adler:2023hxw}.

\begin{figure}[h!]
	\centering
	\includegraphics[width=0.66\columnwidth]{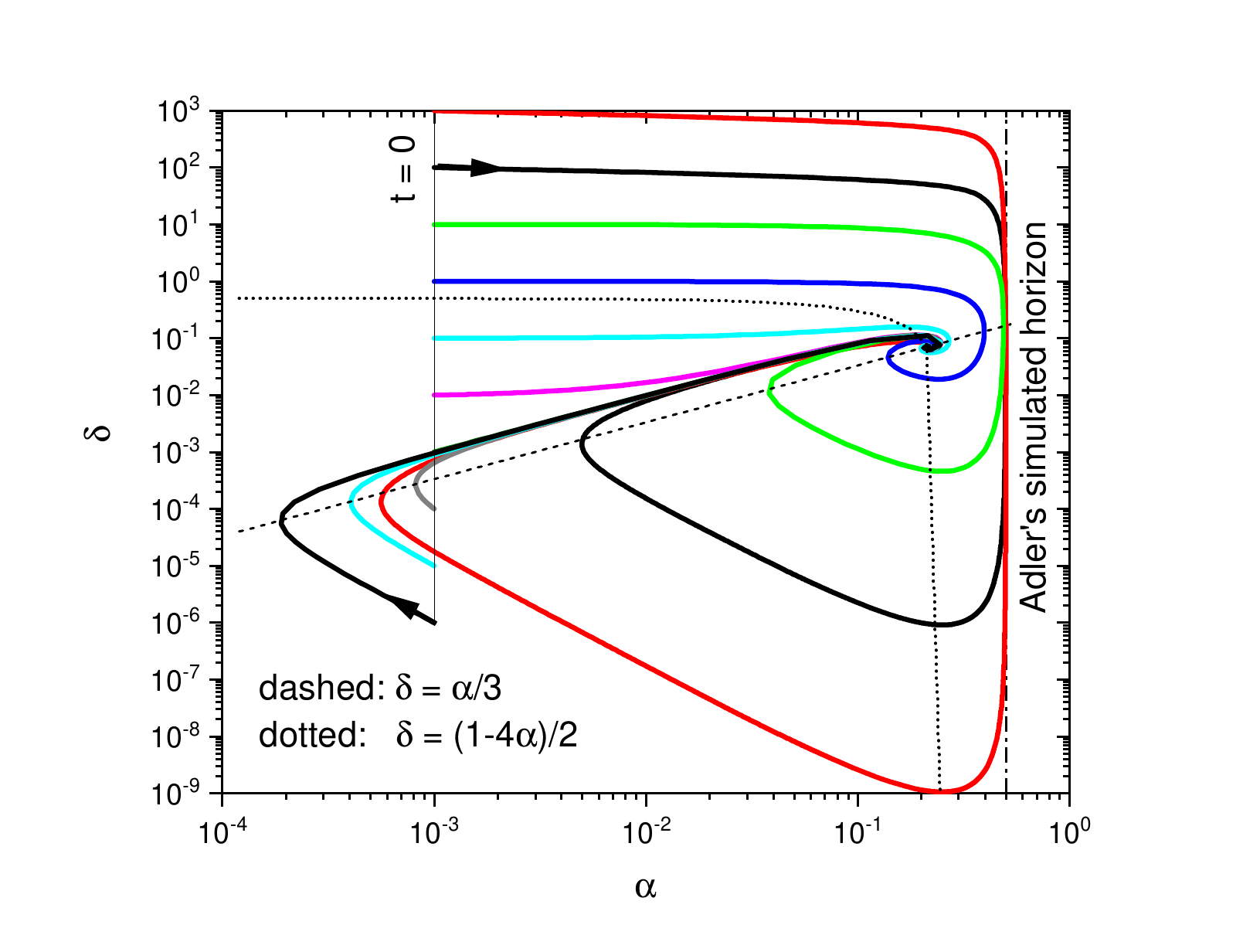}
	\vspace{-3mm}
	\caption{The parametric solutions $\alpha(t)$ and $\delta(t)$ are exhibited as $\delta (\alpha)$
		for $\alpha(t=0) = 10^{-3}$ and  various values of $\delta(t=0)$ (solid curves). 
		The arrows depict the direction of integration of the TOV equations in autonomous form
		which we start at $t = 0$.
		The dashed line is for $\delta = \alpha / 3$, depicting the loci of the extrema of $\alpha (t)$ due to
		$\mathrm{d} \alpha / \mathrm{d} t = 0$, while the dotted curve is for $\delta = (1 - 4 \alpha)/2$,
		depicting the loci of extrema of $\delta (t)$ due to $\mathrm{d} \delta / \mathrm{d} t = 0$.
		The curves spiral into the Misner-Zapolsky point at $\alpha = 3/14$ and $\delta = 1/14$.
		Adler's simulated horizon is for $\alpha \to 0.5$ at $\delta = \alpha/3$.
		Note a second region of squeezed curves ($\delta \approx \alpha$) prior to the Misner-Zapolsky point.
		The curve starting with $\delta(t=0) = 10^{-3}$ is hidden in this bundle.
		\label{fig.1} 
	}
\end{figure}

Figure~\ref{fig.1} exhibits some systematic of the solutions $\alpha (\delta)$ which emerge
from the parametric representations of $\alpha (t)$ and $\delta (t)$.
Starting the integration of Eqs.~(\ref{aTOV1}, \ref{aTOV2}) at $t = 0$ with $t_0 = 0$,
towards positive values of $t$ at selected values of $\alpha (t = 0)$ 
(we chose $10^{-3}$) and $\delta (t  = 0)$,
a group of curves $\delta (\alpha)$ is squeezed in a narrow corridor at $\alpha \to 0.5$,
where in fact $D$ becomes small. Maximum values of $\alpha$ are on the line $\delta = \alpha / 3$
(see dashed line), which depicts the loci of extrema of $\alpha (t)$ 
(or minimum of $D$) due to $\mathrm{d} \alpha / \mathrm{d} t = 0$.
These determine the positions of Adler's kinks or ``simulated horizons''.

The curves, at larger values of $t$, spiral towards the Misner-Zapolsky (MZ) solution~\cite{Misner:1964zz}
at $\alpha = 3/14$ and $\delta = 1/14$, see Fig.~\ref{fig.2}-left below.
In other words, MZ is (besides $\alpha = 0$ and $\delta = 0$) a stable point,
solving the autonomous TOV equations by $\alpha (t) = 3/14$ and $\delta (t) = 1/14$,
explicitly $p(r) = 1 / (56 \pi r^2)$ and $m(r) = (3/14)  r$, i.e.\ a diverging central pressure
and an infinite extent since the pressure drops with $\mathrm{d} p / \mathrm{d} r \propto p$.

Due to the horizontal shift invariance, one can continue the integration into the negative $t$ direction,
i.e.\ opposite to the arrows in Fig.~\ref{fig.1}, meaning inwards integration w.r.t.\ radial coordinate $r$.
In a such a manner, one reaches smaller (or even negative) values of $\alpha$ for the upper curves.
The lower curves are continued, first, to larger values of $\alpha$, then they suffer a back bending and
are squeezed trough the
corridor at $\alpha \to 0.5^-$ with increasing values of $\delta$ and eventually bend to left going again to
lower values of $\alpha$. Thus, they supplement the pattern of curves which start at $\delta (t=0) > (1 - 4 \alpha)/2$
(not displayed).

The pattern of $\delta (\alpha)$ can be obtained also from Bondi's orbit equation
$\mathrm{d} \delta / \mathrm{d} \alpha = \frac{2 \delta (1 - 4 \alpha - 2 \delta)}{(3\delta - \alpha) (1 - 2 \alpha)}$ 
(cf.~\cite{Zurek:1984zz,Collins:1985,Toimela:1986mg} for features and further citations)
by piecemeal  (numerical) integration, either on
$\alpha$ or $\delta$ thus avoiding mutually divergences.
Thereby, the information on $t$ and $r$ increasing or decreasing is lost.  
The orbit in the region $\alpha \ll 1$ and $\delta \gg 1$ reads
$\delta \approx \delta_0 \exp\{ - \frac43 (\alpha - \alpha_0) \}$,
where $\alpha_0$ and $\delta_0$ denote the start points, see upper three curves in Fig.~\ref{fig.1}.  
At $\delta \ll 1$ and $\delta \ll \alpha$, the orbit is about
$\delta \approx \delta_{\min} [\frac{1 - 4 \delta_{\min}^2}{8 \alpha (1 - 2 \alpha)} ]^2$,
where $\delta_{\min}$ is the minimum value at $\alpha (\delta_{\min}) = (1 - 2 \delta_{\min})/4$,
see the three lowest curves with minima at $\alpha \approx 0.25$.
For $\alpha \ll 1$ and $\delta \ll 1$, the orbit reads $\alpha \approx \delta + 2 \delta_\star \sqrt{\delta_\star/\delta }$,
where $\delta_\star$ and $\alpha (\delta_\star) = 3 \delta_\star$ is the position of the respective l.h.s.\ vertex.
The upper (lower) branch is $\delta \approx \alpha$ ($\delta \approx 4 \delta_\star^3 \alpha^{-2}$).
Since the upper branch is independent of $\delta_\star$ in leading order, many curves with vastly different values of
$\delta_\star$ run through the narrow corridor $\delta \approx \alpha$.
Its extension, however, depends on $\delta_\star$: $\alpha > 3 \delta_\star$.

\begin{figure}[h!]
	\includegraphics[width=0.55\columnwidth]{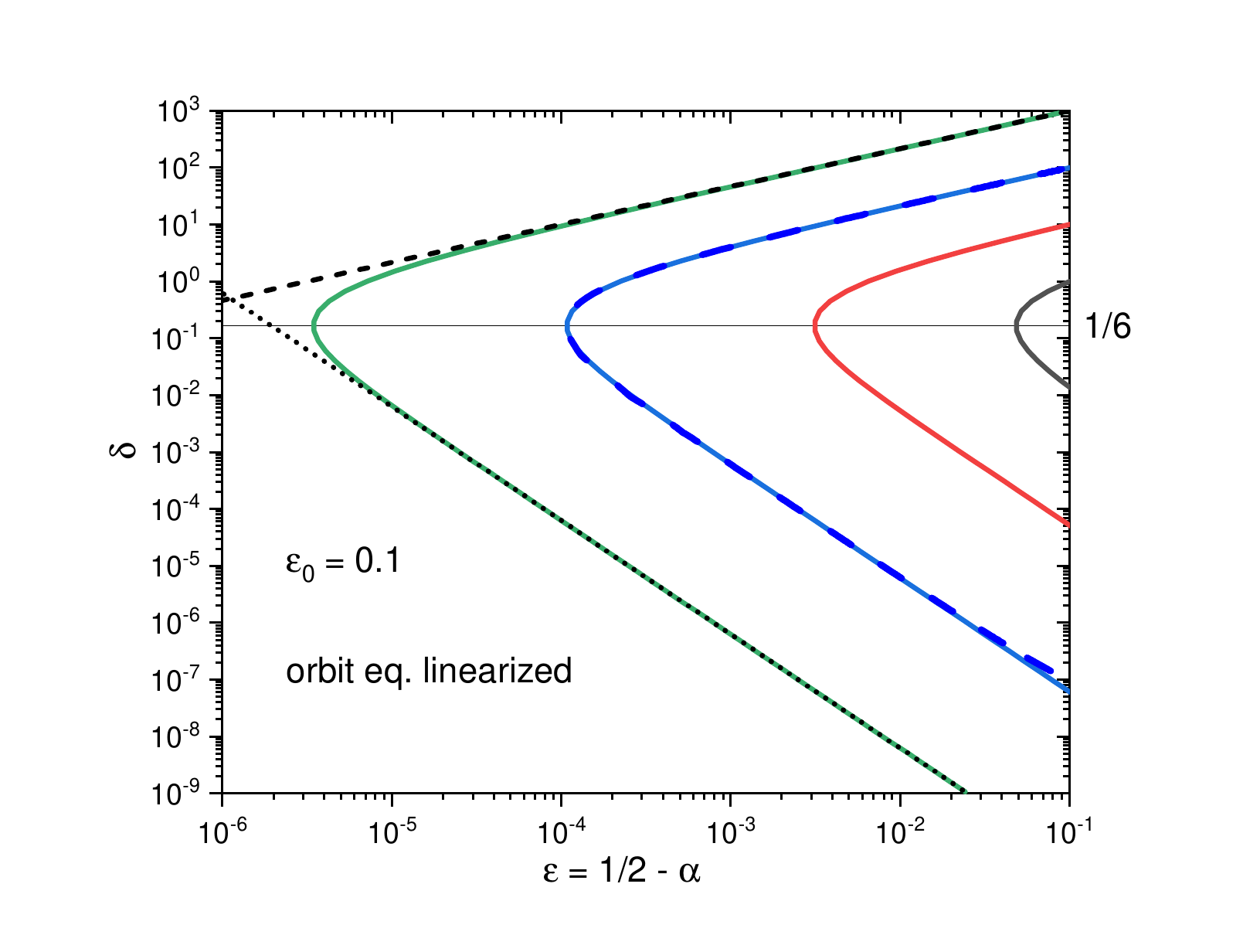} \hspace{-19mm}
	\includegraphics[width=0.55\columnwidth]{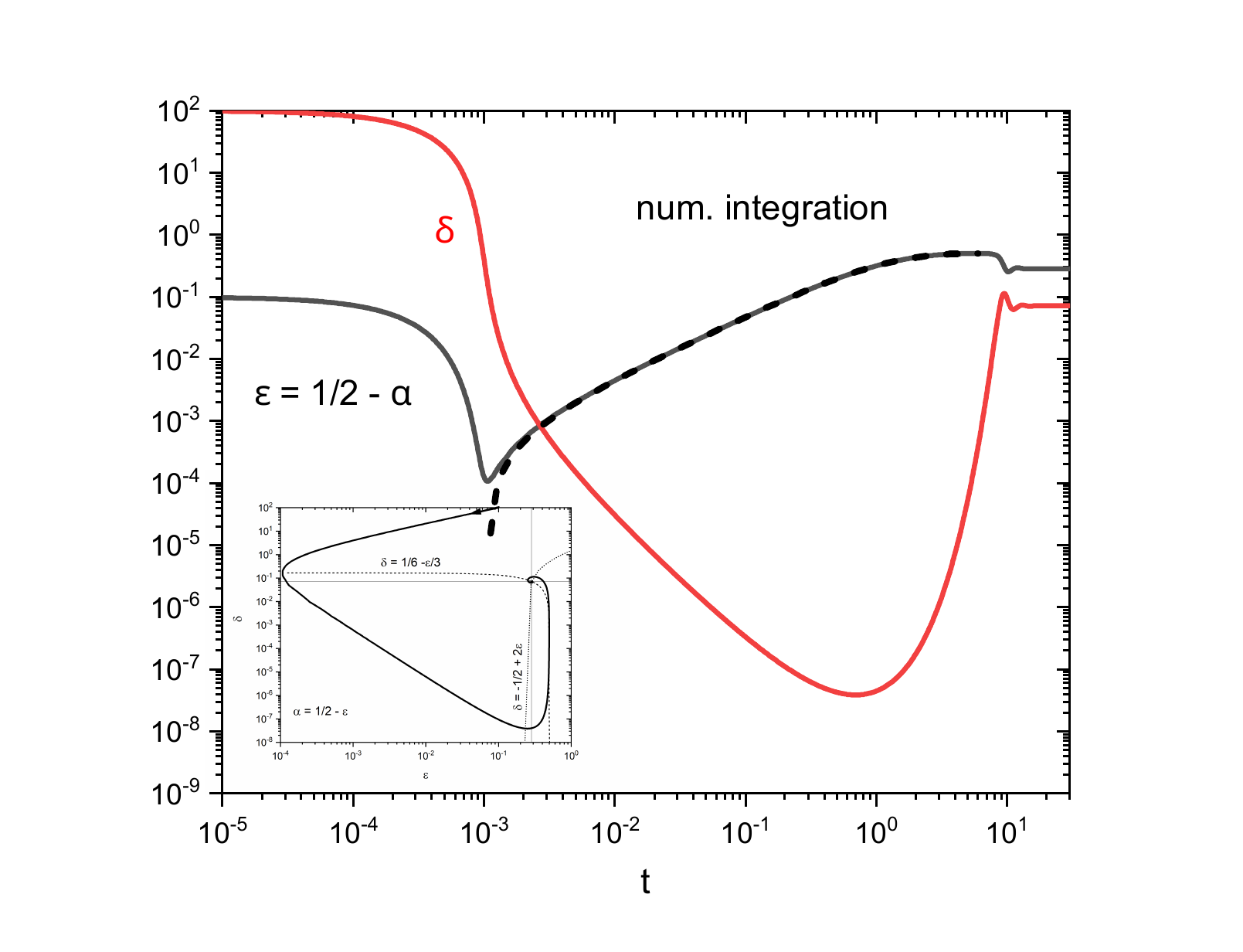} 
	\vspace{-12mm}
	\caption{Left panel: Flipped representation of the quantity $\varepsilon \equiv 1/2 - \alpha =  D / 2$ 
		as a function of $\delta$ for $\varepsilon_0 = 0.1$ and 
		$\delta_0 = 10^0, \ldots, 10^3$ (see intersections of the curves above $\delta = 1/6$ with the r.h.s.\ ordinate)
		from the linearized orbit equation.
		The fat dashed blue curve is from the numerical integration of the autonomous TOV equations
		without approximations.
		The dotted and dashed lines depict the approximations mentioned in the text.
		Right panel: The dependence of $\varepsilon$ (black) and $\delta$ (red) on the variable $t$
		for $\varepsilon_0 = 0.1$ and $\delta_0 = 10^2$.
		%\protect\footnote{
			%Choosing initial values $\varepsilon_0 = 0.1$ and $\delta_0 = 2 \times 10^2$ 
			%reproduces $D(t)= 2 \varepsilon(t)$ of Fig.~4 in \cite{Adler:2023hxw} at $t_\mathrm{Adler} > 0.0017$
			%with a suitable shift of $t$. }
		Adler's kink is visible at $t_{\mathrm{kink}} \approx 1.087 \times 10^{-3}$.
		It corresponds to the l.h.s.\ vertex of the blue curve in the left panel.
		The dashed curve depicts $\varepsilon (t) \approx (1 - \exp \{ t_{\mathrm{kink}} - t \} )/2$ for $t_{\mathrm{kink}} < t < 6.66$.
		The tiny inset shows how $\varepsilon (t)$  and $\delta(t)$ combine to the orbit $\delta (\varepsilon)$, 
		thus complementing the blue curves in the left panel. Loci of the extrema of $\varepsilon$ and $\delta$ are indicated 
		by dashed ($\delta = 1/6 - \varepsilon /3$) and dotted ($\delta = - 1/2 + 2 \varepsilon$) curves.
		The MZ point is at the crossing of the thin lines.
		\label{fig.1b} 
	}
\end{figure}

To decipher the pattern of curves near Adler's simulated horizon, i.e.\ at $\alpha \to 1/2$ and 
$\delta = \mathcal{O}(0.1)$, it is instructive to turn to the suitable variable 
$\varepsilon := \frac12 - \alpha = \frac12 D$.
The linearized orbit equation for $\varepsilon \to 0$ becomes then
$\mathrm{d} \delta / \mathrm{d} \varepsilon = \frac{2 \delta (1 + 2 \delta)}{\varepsilon (- 1 + 6 \delta)}$,
which can be integrated to get
$\varepsilon = \varepsilon_0 \left( \frac{1 + 2 \delta}{1 + 2 \delta_0} \right)^2 \sqrt{\frac{\delta_0}{\delta}}$.
Figure \ref{fig.1b}-left exhibits $\delta(\varepsilon)$. 
The extremum, $\mathrm{d} \varepsilon / \mathrm{d}\delta = 0$, is found at $\delta_{\min} \approx 1/6$ with
$\varepsilon (\delta_{\min}, \delta_0 \gg 1) \approx \sqrt{32/27} \varepsilon_0 \delta_0^{- 3/2}$, being the positions
of the l.h.s.\ vertices. That means that,
with increasing values of $\delta_0$ at fixed $\varepsilon_0 \ll 1$, one gets smaller minimum values of
$\varepsilon$ or $D$, i.e.\ Adler's kink becomes deeper,
$\propto \delta_0^{-3/2}$.
Note the scaling $\propto \varepsilon_0$.
For $\delta, \delta_0 \gg 1$ ($\ll 1$) we find the useful approximations
$\varepsilon \approx \varepsilon_0 (\delta / \delta_0)^{3/2}$ (dashed line), respectively
$\varepsilon \approx \frac{\varepsilon_0}{4 \delta_0^{3/2}} \delta^{-1/2}$ (dotted line),
both displayed for $\varepsilon_0 = 0.1$ and $\delta_0 = 10^3$. 
Converting $\varepsilon \to \alpha$ explains the bundling of curves towards $\alpha = 1/2$ around
$\delta = 1/6$ seen in Fig.~\ref{fig.1}.

Figure \ref{fig.1b}-right exhibits the $t$ dependence of $\varepsilon$ (black curve) and $\delta$ (red curve).
Adler's kink is located at $t \approx 10^{-3}$. The tiny up's and down's at $t > 7$ 
signal the approach into Collins' spiral focus (see inset).
The quantity $m(r) = \alpha(r(t)) r$ ($p(r) = \delta(r(t)) /(4 \pi r^2)$) with $r = r_0 \exp\{ t \}$
is continuously increasing (decreasing), irrespective of $r_0$,
despite the strongly decreasing (increasing) sections of $\alpha (t)$ ($\delta (t)$). 
In particular, $m(r(t))$ is nearly flat in the interval $t \in [t_{\mathrm{kink}}, 6.66]$.
There, the approximation $\varepsilon (t) \approx (1 - \exp \{ t_{\mathrm{kink}} - t \} )/2$ (dashed curve) applies.
The dashed curve is nothing than
an approximation of the outer Schwarzschild solution $\propto 1 - const / r$ with suitable $const$
in some appropriate interval of the radius coordinate. 
This is at the origin of Adler's ``simulated horizon''~\cite{Adler:2023hxw}.

\begin{figure}[h!]
	\includegraphics[width=0.49\columnwidth]{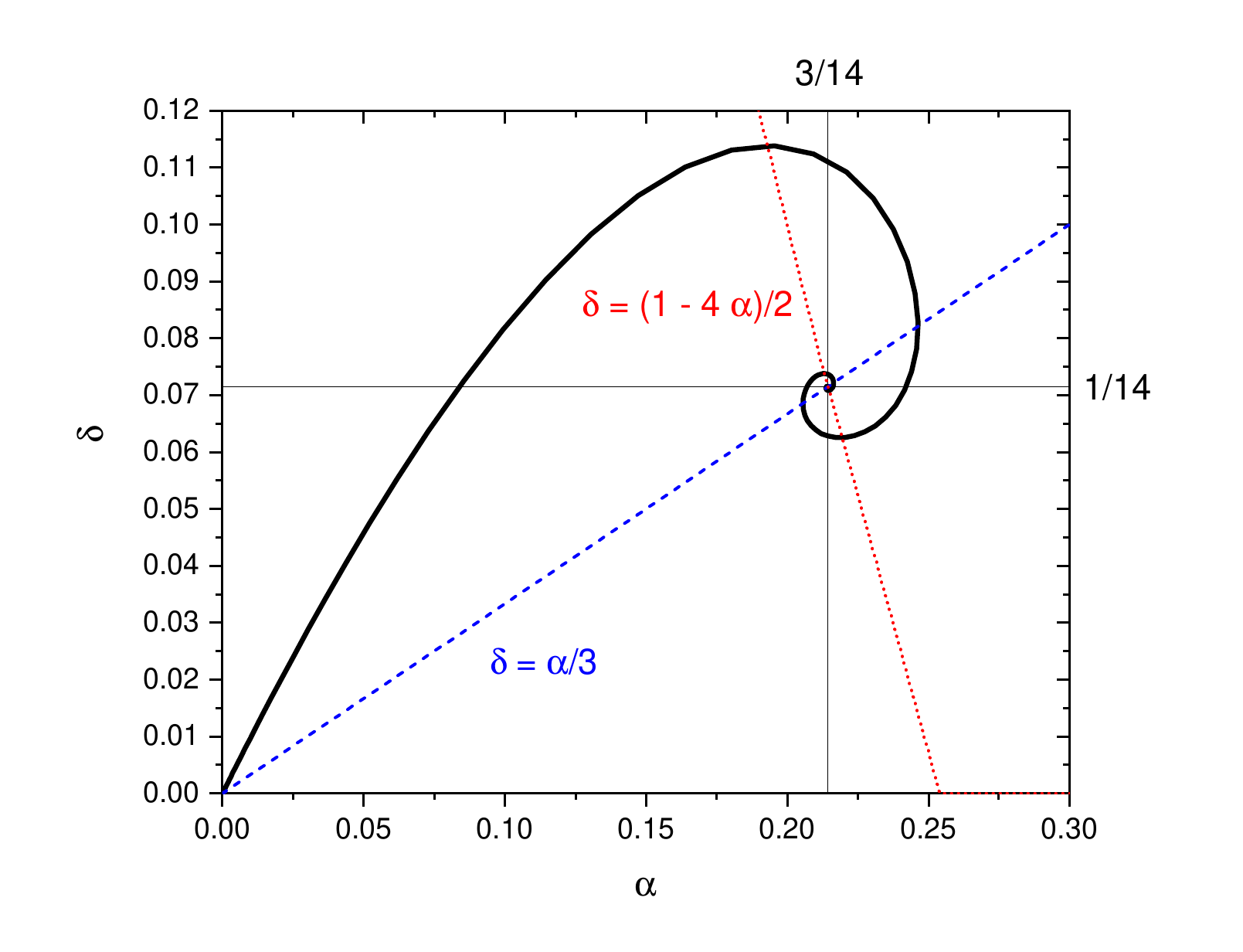}
	\includegraphics[width=0.49\columnwidth]{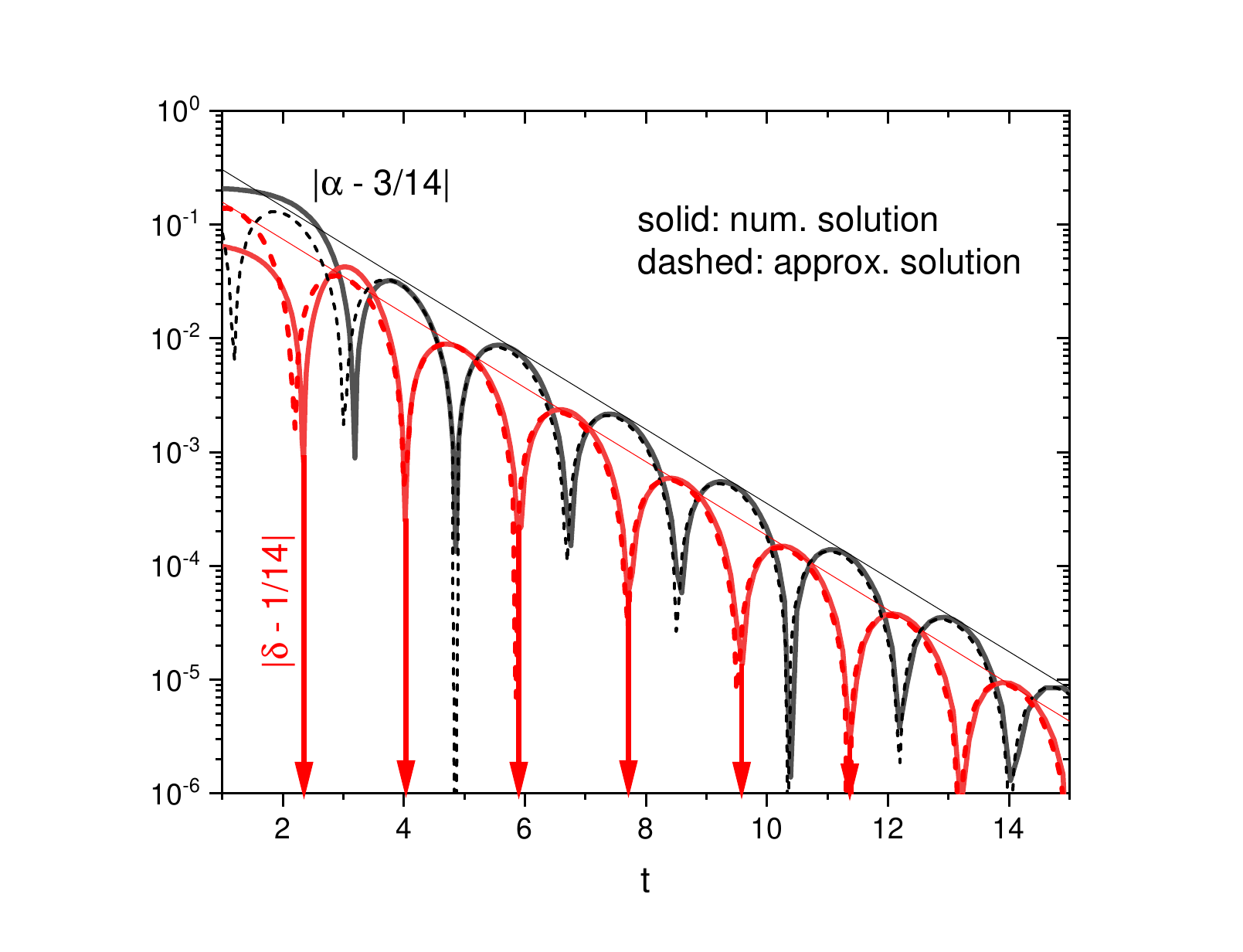}
	\vspace{-3mm}
	\caption{Left panel: As Fig.~\ref{fig.1} but in linear representation and a zoom to expose the spiraling \cite{Collins:1985}
		into the Misner-Zapolsky point. For $\alpha(t=0) = \delta (t=0) = 10^{-3}$. 
		Right panel: The corresponding dependence of $A := \alpha - \frac{3}{14}$ (black curve) and 
		$B := \delta - \frac{1}{14}$ (red curve) on $t$. The modulus of $A$ and $B$ is displayed
		to cope with the log scale.
		The dashed curves depict the approximations based on Eq.~(\ref{Approx}).
		%$A \approx c_1 \exp\{ - \frac{3}{4} (t - \hat t_0) \}  \sin \left\{ \frac{\sqrt{47}}{4} (t - \hat t_0) \right\}$
		%with $c_1 = 0.02$ and $\hat t_0 = 4.85$. 
		The thin lines exhibit the decays  $\propto  \exp\{ - \frac{3}{4} (t - \hat t_0) \}$ with proper normalizations.
		%($c_1$ for the black line and $0.0078$ for the red line).
		\label{fig.2} 
	}
\end{figure}

Figure \ref{fig.2}-left displays Collins' spiral, for $\alpha(t=0) = \delta(t=0) = 10^{-3}$, in a linear scale. 
%(Other curves for $\delta(t=0) < 10^{-2}$ are on top.)
To quantify further the approach to the MZ point, Fig.~\ref{fig.2}-right exhibits the modulus of the shifted quantities
of $A := \alpha - \frac{3}{14}$ (black curve) and 
$B := \delta - \frac{1}{14}$ (red curves). 
The crossings of the respective zeroes are marked by red arrows for $B$ only.
The above autonomous TOV equations (\ref{aTOV1}, \ref{aTOV2}) can be converted into a second-derivative
equation for $A$\footnote{
	Intermediate steps are
	$\dot B = -\frac{1 + 7 B}{2 - 7 A}(2A + B)$ and insertion of $B = \frac13 (\dot A - A)$,
	respectively $\dot B = \frac13 (\ddot A - \dot A)$.
	The derivative w.r.t.\ $t$ is denoted by a dot.},
which can be linearized, for small $A$ and $B$: $\ddot A + \frac32 \dot A + \frac72 A = 0$ 
with the solution
\begin{align} \label{Approx}
	A \approx \exp\{- \tfrac{3}{4} (t - \hat t_0) \} 
	[ c_1 \sin \{ \tfrac{\sqrt{47}}{4} (t - \hat t_0) \}
	+  c_2 \cos \{ \tfrac{\sqrt{47}}{4} (t - \hat t_0) \} ] . 
\end{align}
The choice $c_1 = 0.02$, $c_2 = 0$ and $\hat t_0 = 4.85$ generates the black dashed curve.
This approximation is used to derive the corresponding approximation $B \approx (\dot A + A)/3$
(red dashed curve).
Despite the linearization, the numerical solution is fairly well reproduced
in the region where $A$ and $B$ are small enough.
The slopes of the decay of $A$ and $B$ are also well described (see black and red thin lines).  
The distance of the spiral to its center, $\sqrt{A^2 + B^2}$, as a function of $t$, wobbles
in the corridor $(0.33 \ldots 0.7) \exp \{- \frac34 t \}$ for $t > 3$ (not displayed). 
The function $m(r)$ ($p(r)$) is continuously increasing (decreasing) despite the oscillations of 
$\alpha (t)$ and $A(t)$ ($\delta (t)$ and $B(t)$).

Introducing a scale in the EoS, e.g.\ $e = v_s^{-2} p + e_s$, where $e_s$ stands for the energy density
of the configuration's surface at zero pressure, the scale invariance is lost:
$\mathrm{d} \alpha / \mathrm{d} t = \delta \frac{e}{p} - \alpha$,
$\mathrm{d} \delta / \mathrm{d} t = 2 \delta  - \delta (1 + \frac{e}{p})(\alpha + \delta)/ (1- 2 \alpha)$,
${e}/{p} := 3 + \frac{e_s}{p_0} \frac{\delta_0}{\delta} \exp\{ 2 (t - t_0) \}$.
The orbits terminate by steep dropping $\delta \to 0$ prior to the MZ point, depending sensitively on $\frac{e_s}{p_0}$.

\end{appendices}
	\bibliography{MDPI.bib}
	\bibliographystyle{tip2.bst}
\end{document}